\documentclass[pra, reprint, twocolumn, superscriptaddress, amsfonts, amssymb, amsmath]{revtex4-2}
\usepackage[english]{babel}

\usepackage{graphicx}

\graphicspath{{./Images/},{./ImagesAppendix/},
{./images/},{./imagesAppendix/}}

\usepackage{physics}
\usepackage{bm}
\usepackage{braket}
\usepackage{pgfplots}
\usepackage{array}
\usepackage{makecell}
\usepackage{amsmath}
\usepackage{tikz}
\usetikzlibrary{positioning, arrows.meta}
\usepackage{subfigure}
\usepackage{commath}
\usepackage{graphicx,bm}
\usepackage{verbatim}
\usepackage{flafter}
\usepackage{float}
\usepackage{bbold}
\usepackage{varioref}

\usepackage{mathtools}

\DeclarePairedDelimiter\floor{\lfloor}{\rfloor}

\usepackage{color}
\usepackage[colorlinks,citecolor=darkBlue,linkcolor=darkBlue,
urlcolor=blue,hyperindex]{hyperref}
\definecolor{darkBlue}{rgb}{0.08, 0.13, 0.4}

\definecolor{THc}{rgb}{0.9,0.3,0.2}

\begin{document}

\title{Extractable energy from quantum superposition of current states}
\author{Francesco Perciavalle}
\affiliation{Quantum Research Center, Technology Innovation Institute, P.O. Box 9639 Abu Dhabi, UAE}
\affiliation{Dipartimento di Fisica dell’Universit\`a di Pisa and INFN, Largo Pontecorvo 3, I-56127 Pisa, Italy}

\author{Davide Rossini}
\affiliation{Dipartimento di Fisica dell’Universit\`a di Pisa and INFN, Largo Pontecorvo 3, I-56127 Pisa, Italy}

\author{Juan Polo}
\affiliation{Quantum Research Center, Technology Innovation Institute, P.O. Box 9639 Abu Dhabi, UAE}

\author{Luigi Amico}
\affiliation{Quantum Research Center, Technology Innovation Institute, P.O. Box 9639 Abu Dhabi, UAE}
\address{Dipartimento di Fisica e Astronomia ``Ettore Majorana" University of Catania, Via S. Sofia 64, 95123 Catania, Italy}
\affiliation{INFN-Sezione di Catania, Via S. Sofia 64, 95123 Catania, Italy}

\date{\today}

\begin{abstract}
We explore the energy content of superpositions of current states. Specifically, we focus on the maximum energy that can be extracted from them  through local unitary transformations. The figure of merit we employ is the local ergotropy. We perform a complete analysis  in the whole range of the system's parameters. 
This way, we prove that superpositions of two current states in spatially closed spin networks are characterized by specific peaks in extractable energy, generally overcoming the ergotropy of each of the two separate current states characterized by a single winding number. The many-body state dynamics entails to  ergotropy evolving in a controlled fashion. The implementation we suggest is based on a Rydberg-atom platform. 
Optimal transformations able to extract locally the maximum possible amount of energy are sorted out. 
\end{abstract}

\maketitle

\section{Introduction}
Quantum superposition states underpin the second quantum revolution that is focused on quantum technology. They have been proven to speedup  computation protocols, enhance the sensitivity of quantum devices, and empower secure communication protocols. 
A relevant example is given by the fundamental computational unit of quantum computing, known as the qubit, which is a superposition of two distinct logical states~\cite{nielsen2012quantum}. The manipulation of quantum superposition states lies at the heart of quantum technology~\cite{dowling2003quantum} and can lead to entanglement in extended many-body quantum systems~\cite{amico2008entanglement,horodeck2009quantum,leggett2002testing}. The quantum superposition principle  plays  a key role to develop quantum sensors with enhanced precision and stability~\cite{szigeti2021improving,degen2017quantum,giovannetti2011advances}, to devise secure communication schemes~\cite{gisin2007quantum}, and to reach quantum advantage
in quantum simulation and computation~\cite{nielsen2012quantum,grover1996fast}.  In the latter context, among the others, we highlight how  Rydberg-atom platform can provide quantum gates with a feasible scalability and controllability~\cite{adams2019rydberg,jaksch2000fast,bluvstein2023logical,barredo2016atom}. 

In the present paper, we consider superposition of current states. Such states can be implemented  in superconducting circuits~\cite{leggett2002testing, vanderWal2000quantum, orlando1999superconducting, clarke2008superconducting} and atomtronic circuits~\cite{amico2022colloquim, aghamalyan2015coherent}. Here, we focus on superpositions of spin-waves. Such states have been theoretically proposed to engineer chiral flows of excitations in networks of Rydberg atoms~\cite{perciavalle2024quantum}. Among different features, such states are demonstrated to enjoy a specifically controllable time evolution   that may be exploited for precise manipulations  of Rydberg excitations in a quantum network. Chiral flows have been studied also in~\cite{debernardis2021light,lienhard2020realization, wang2023chiral,roushan2016chiral,lu2016chiral,valencia2024rydberg} with the aim of achieving controlled transport of energy and information~\cite{palaiodimopoulos2024chiral, zimbors2013quantum,bottarelli2023quantum}. 
We note that, while such flows involve both spin and energy transfer, these two notions can be distinct~\cite{popkov2013manipulating,xu2019transport, balachandran2019heat, mendozaArenas2013heat}. 

\begin{figure}[!t]
\centering
\includegraphics[width=1\columnwidth]{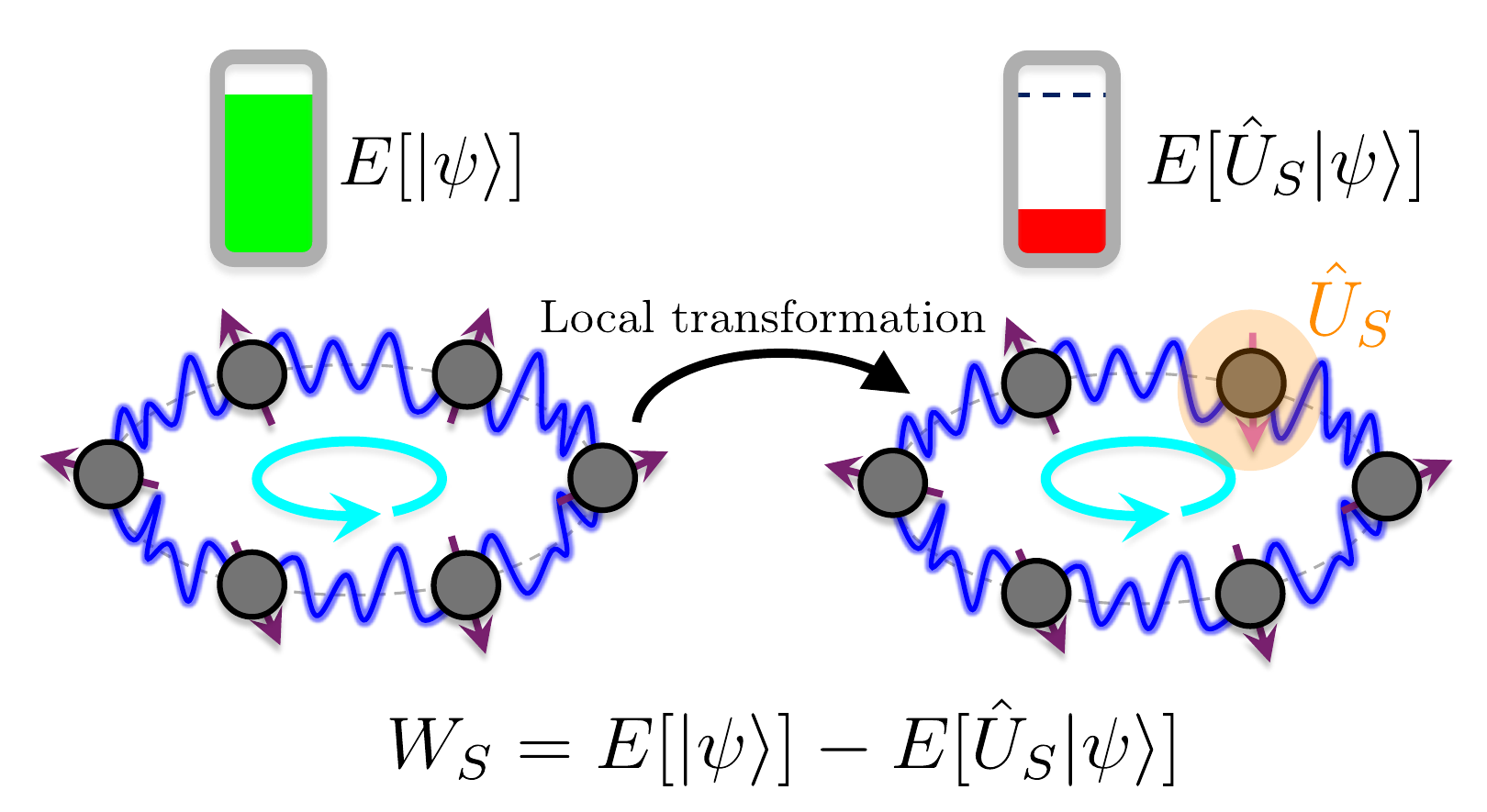}
\caption{Schematic representation of the energy extraction process. An interacting ring-shaped spin system has an initial energy $E[\ket{\psi}]$, the site $S$ is subject to a local rotation $\hat{U}_S$, the energy after the rotation is $E[\hat{U}_S\ket{\psi}]$. If positive, the difference between initial and final energies $W_S$ gives the extractable energy.}
\label{fig:Sketch}
\end{figure}

We specifically focus on the study of the flow of energy encoded in  currents of Rydberg excitations. 
We ask questions like:  How can we characterize such spin current states from an energetic perspective? How can we quantify the extractable   energy from the system's state? And finally: Does  quantum superposition of currents states bring an energetic adavantage? 
To face these questions, we consider a ring-shaped interacting  spin system 
and quantify the  amount of energy that can be extracted from the system's quantum states through the notion of {\em ergotropy}~\cite{allahverdyan2004maximal}. Originally introduced in quantum thermodynamics~\cite{vinjanampathy2016quantum, alicki1979quantum}, the global ergotropy is a figure of merit for the energy extracted through generic non-local unitary operations. Recent studies have shown its usefulness in the context of quantum technologies, to quantify the extractable work in devices aimed at storing energy, where peculiar quantum properties like entanglement may provide a net advantage over classical analogues~\cite{alicki2013entanglement,campaioli2024colloquim, binder2015quantacell,campaioli2017enhancing,rossini2020quantum, andolina2019quantum,rossini219many,catalano2024frustrating,manzano2018optimal, gyhm2024beneficial, konar2022quantum, konar2024quantum, simon2024correlations}. 
Since we are interested on how the energy distributes in a closed system, and  relying on the local controllability that Rydberg atoms platforms enjoy~\cite{bornet2024enhancing}, we specifically refer to the \textit{Local Ergotropy} (LE), namely the maximum extractable energy when applying local unitaries~\cite{salvia2023optimal,mukherjee2016presence,imai2023work, castellano2024extended, dibello2024locale}. 

In this paper, we employ the LE as a figure of merit characterizing the  energetics of the class of states of interest in the whole range of system's parameters. 
The study of the dynamics demonstrates a chiral flow of extractable energy along the ring that does not spread unevenly through the system. We identify the optimal transformations required to extract the maximum possible amount of energy. We find that quantum superpositions of current states can bring an energetic advantage with respect to single currents, since superposition states lead to localized ergotropy in specific areas of the system, on demand. 

The paper is organized as follows. In Sec.~\ref{sec:ergo}, we introduce the concept of ergotropy, highlighting the LE and the closed formula to compute it. In Sec.~\ref{sec:local_ergo_deltaJ}, we study the LE of a quantum superposition of two current states, analyze its distribution, and compare it with those of single current states, showing that in our specific system optimal or quasi-optimal transformations are simple operations. In Sec.~\ref{sec:chiral_ergo}, we study the dynamics comparing the controlled LE evolution of the superposition of current states with those of another state made of a Bell state in two specific sites of the ring. In Sec.~\ref{sec:ergo_dipolarXY}, we show how to compute the LE in presence of long-range hopping, focusing on the dipolar case corresponding to Rydberg-atom systems. Finally, in Sec.~\ref{sec:outlook}, we summarize our results and draw some possible future perspectives.

\section{Global and local ergotropy of an interacting spin model}
\label{sec:ergo}

The ergotropy is a physical quantity introduced in quantum thermodynamics~\cite{vinjanampathy2016quantum} to estimate the maximum extractable work from a quantum system~\cite{allahverdyan2004maximal}. Given a quantum system in a state $\rho$ living in a Hilbert space $\mathcal{H}$ described by a Hamiltonian $\hat{H}$, the ergotropy is defined as~\cite{allahverdyan2004maximal,alicki2013entanglement,campaioli2024colloquim}
\begin{equation}
\mathcal{E}\bigl(\rho,\hat{H}\bigr)=\max_{\hat{U}} \Tr \left[ \hat{H}\rho - \hat{H} \hat{U}\rho \hat{U}^{\dagger}\right].    
\end{equation}
Here $\mathcal{E}\bigl(\rho,\hat{H}\bigr)$ denotes the global or the local ergotropy, depending on whether $\hat{U}$ is a global or a local unitary operator, respectively.
In this work, we specifically refer to the local case~\cite{salvia2023optimal}. Specifically, given a bipartite system in a Hilbert space
$\mathcal{H}=\mathcal{H}_S \otimes \mathcal{H}_E$, where $S$ is the subsystem on which the unitary operations are performed, and $E$ indicates the environment, the global Hamiltonian can be written as
\begin{equation}
    \hat{H}=\hat{H}_S \otimes \hat{\mathbb{1}}_E + \hat{\mathbb{1}}_S \otimes \hat{H}_E + \hat{V}_{SE},
\end{equation}
the maximum extractable work from the subsystem $S$, meaning the $S$-LE is
\begin{equation}
\mathcal{E}_S(\rho_{SE},\hat{H}) = \max_{\hat{U}_S} \Tr \! \left[\hat{H}\rho_{SE} \! - \! \hat{H} (\hat{U}_S \! \otimes \! \hat{\mathbb{1}}_E)\rho_{SE} (\hat{U}_S^{\dagger} \! \otimes \! \hat{\mathbb{1}}_E)\right]
\end{equation}
$\rho_{SE}$ being the state of the whole system $S+E$ in the full space $\mathcal{H}$. If the latter is a pure state $\ket{\psi}$, the LE can be written as
\begin{equation}
\mathcal{E}_S(\ket{\psi},\hat{H}) = \max_{\hat{U}_S} [W_{S}],
\quad
W_S = E[\ket{\psi}] - E[\hat{U}_S \ket{\psi}],
\label{eqref:optimal}
\end{equation}
where $E[\ket{\psi}]=\braket{\psi|\hat{H}|\psi}$ is the mean energy of the pure state $\ket{\psi}$ and $W_S$ is the extractable work through a local rotation $\hat{U}_S$. The energy extraction process is sketched in Fig.~\ref{fig:Sketch}, where the local unitary is performed in a ring-shaped interacting spin system. If the energy of the system is decreased after the local operation, the difference between initial and final energies can be identified as the extractable energy.

We consider the case in which the subsystem $S$ is a single qubit. The LE can be computed using a closed formula obtained in Ref.~\cite{salvia2023optimal}. The computation is carried out  by introducing the $3\times 3$ matrix $\mathcal{M}$
\begin{equation}
    \mathcal{M}_{jk}=-\Big( r_j h_k + \tfrac{1}{2} {\rm Tr} \big\{ \rho_E^{(j)}\hat{V}_E^{(k)} \big\} \Big), \quad j,k \!=\! \{x,y,z\},
    \label{eq:Mmatr_def}
\end{equation}
where 
\begin{align}
    r_j = \Tr \big\{ \hat{\sigma}_S^{(j)}\rho_S \big\} , \quad &  \hat{V}_{E}^{(k)} = \Tr_S\{\hat{\sigma}_S^{(k)}\hat{V}_{SE}\}, \nonumber \\
    h_k = \tfrac12 \Tr \big\{ \hat{\sigma}_S^{(k)}\hat{H}_S \big\} , \quad & \rho_E^{(j)} = \Tr_S\{\hat{\sigma}_S^{(j)}\rho_{SE}\},
    \label{eq:Mmatr_def2}
\end{align}
with $\rho_S=\Tr_E \{\rho_{SE} \}$ being the reduced state in the subsystem subspace. The LE reads~\cite{salvia2023optimal}
\begin{equation}
\mathcal{E}_S(\rho_{SE},\hat{H}) \! = \!
\begin{cases}
\Tr{|\mathcal{M}|-\mathcal{M}} & \!\! \rm{if} \; \rm{Det}[\mathcal{M}] \! \geq \! 0, \\
\Tr{|\mathcal{M}|-\mathcal{M}} \!-\! \dfrac{2}{\lVert \mathcal{M}^{-1} \rVert} & \!\! \rm{if} \; \rm{Det}[\mathcal{M}] \! < \! 0,
\end{cases}
\end{equation}
where $|\mathcal{M}|=\sqrt{\mathcal{M}^{\dagger}\mathcal{M}}$ and ${\lVert \mathcal{M}^{-1} \rVert}$ is the operator norm of the inverse of the $\mathcal{M}$ matrix.

\subsection{Local ergotropy of the XY model in the single-excitation sector}

We consider a ring composed of $L$ qubits described by the so-called XY spin-chain model:
\begin{equation}
  \hat{H} = J \sum_j \big( \hat{\sigma}_j^x\hat{\sigma}_{j+1}^x + \hat{\sigma}_j^y\hat{\sigma}_{j+1}^y \big) + \Delta \sum_j \hat{\sigma}_j^z,
  \label{eq:XY_NN_ham}
\end{equation}
where $\hat \sigma^\alpha_j$ are the spin-1/2 Pauli matrices ($\alpha=x,y,z$), $J$ is spin exchange and $\Delta$ is a transverse field. 
In Eq.~\eqref{eq:XY_NN_ham}, we adopt periodic boundary conditions, by assuming $\hat \sigma^\alpha_{L+1} = \hat \sigma^\alpha_1$. Hereafter, unless specified, all summations are assumed to run from $1$ to $L$.
The subsystem, the environment, and the system-environment coupling Hamiltonians are respectively given by
\begin{subequations}
\label{eq:XYmodel}
\begin{eqnarray}
\hat{H}_S & = & \Delta \hat{\sigma}_S^z, 
\label{eq:Ham_S}\\
\hat{H}_E & = & \Delta \sum_{j \neq S} \hat{\sigma}_j^z + J\sum_{j\neq S,S-1} \big( \hat{\sigma}_j^x\hat{\sigma}_{j+1}^x + \hat{\sigma}_j^y\hat{\sigma}_{j+1}^y \big), \\
\hat{V}_{SE} & = & J \big( \hat{\sigma}^x_{S}\hat{\sigma}^x_{S+1} + \hat{\sigma}^x_{S}\hat{\sigma}^x_{S-1} + \hat{\sigma}^y_{S}\hat{\sigma}^y_{S+1} + \hat{\sigma}^y_{S}\hat{\sigma}^y_{S-1} \big). \qquad
\end{eqnarray}
\end{subequations}
As explicitly shown in Appendix~\ref{app:M_XYmodelNN}, the matrix $\mathcal{M}$ in Eq.~\eqref{eq:Mmatr_def} can be rewritten as~\cite{salvia2023optimal}
\begin{equation}
\mathcal{M} \! = \! - \!
\begin{bmatrix}
J\braket{\hat{\sigma}_S^x \otimes \hat{X}_E}_{\psi} & J\braket{\hat{\sigma}_S^x \otimes \hat{Y}_E}_{\psi}  & \Delta\braket{\hat{\sigma}_S^x}_{\psi}\\
J\braket{\hat{\sigma}_S^y \otimes \hat{X}_E}_{\psi}  & J\braket{\hat{\sigma}_S^y \otimes \hat{Y}_E}_{\psi}  & \Delta\braket{\hat{\sigma}_S^y}_{\psi} \\
J\braket{\hat{\sigma}_S^z \otimes \hat{X}_E}_{\psi}  & J\braket{\hat{\sigma}_S^z \otimes \hat{Y}_E}_{\psi}  & \Delta\braket{\hat{\sigma}_S^z}_{\psi}
\end{bmatrix} \!, 
\label{eq:M_matrix}
\end{equation}
where $\hat{X}_E = \hat{\sigma}_{S-1}^x + \hat{\sigma}^x_{S+1}$, $\hat{Y}_E = \hat{\sigma}_{S-1}^y + \hat{\sigma}^y_{S+1}$, and $\braket{\ldots }_{\psi}=\braket{\psi|...|\psi}$ indicates expectation values on the state $\rho_{SE}=\ket{\psi}\bra{\psi}$, with $\ket{\psi} \in \mathcal{H}_S \otimes \mathcal{H}_E$.

Let us now focus on the one-excitation sector (1s), i.e., in the space spanned by wave vectors of the form
\begin{equation}
    \ket{\psi_{\rm 1s}} = \sum_j f_j \ket{j}, \qquad\ket{j} = \hat \sigma^+_j \ket{\downarrow, \ldots, \downarrow},
    \label{eq:psi_1s}
\end{equation}
where $\sum_j |f_j|^2 = 1$ ensures the proper normalization. For that class of states, one can show that
$\mathcal{M}_{zx}=\mathcal{M}_{zy}=\mathcal{M}_{xz}=\mathcal{M}_{yz}=0$. Moreover, the relations
$\mathcal{M}_{xx}= \mathcal{M}_{yy}$ and
$\mathcal{M}_{xy}= -\mathcal{M}_{yx}$
hold, since $\hat{\sigma}_i^x\hat{\sigma}_j^x=\hat{\sigma}_i^y\hat{\sigma}_j^y$ and $\hat{\sigma}_i^x\hat{\sigma}_j^y=-\hat{\sigma}_i^y\hat{\sigma}_j^x$ (see Appendix~\ref{app:M_XYmodelNN}). Then we obtain: 
\begin{eqnarray}
\mathcal{E}_S\!\left(\mathcal{M}_{zz} \! \geq \! 0\right) & = & 2 \Big[ \sqrt{\mathcal{M}_{xx}^2 \!+\! \mathcal{M}_{xy}^2} \!  -\! \mathcal{M}_{xx} \Big], 
\label{eq:loc_erg} \\
\mathcal{E}_S\!\left(\mathcal{M}_{zz} \! < \! 0\right) & = & 
2 \Big[ \max \! \Big\{ \sqrt{\mathcal{M}_{xx}^2 \!+\! \mathcal{M}_{xy}^2} , |\mathcal{M}_{zz}| \Big\} \! - \! \mathcal{M}_{xx} \Big]. \nonumber
\end{eqnarray}
We note that the LE depends linearly on the field $\Delta$ only for specific values of the latter.

\begin{figure}[!h]
\centering
\includegraphics[width=\linewidth]{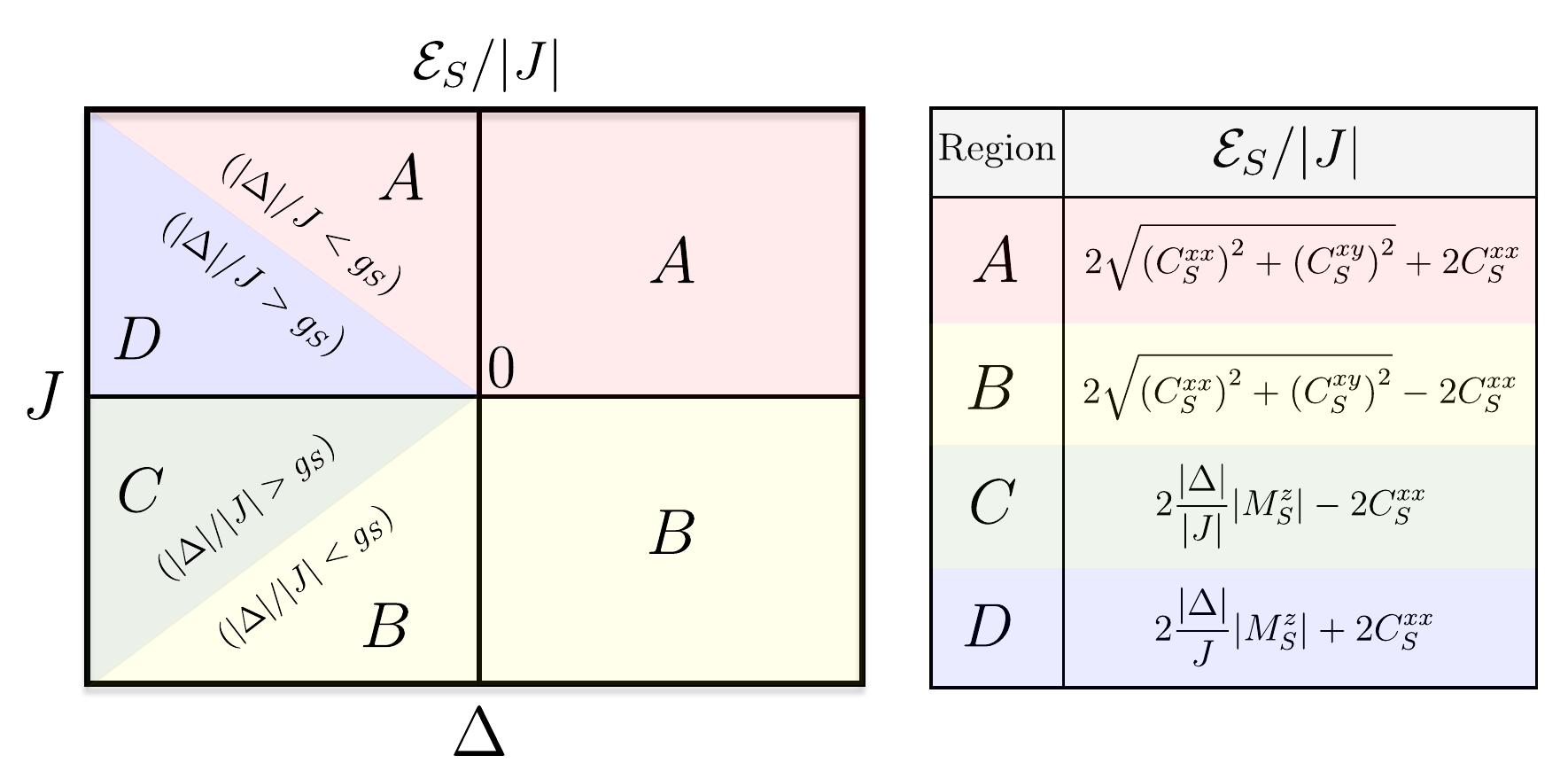}
\caption{Sketch of the LE behavior in the $(\Delta,J)$ plane. The table on the right shows the mathematical expressions of the LE in the different colored regions of the plot on the left.}
\label{fig:sketch_ergotropy}
\end{figure}

\section{Local ergotropy of the superposition of current states}
\label{sec:local_ergo_deltaJ}

We consider a particular class of quantum states which consists of superpositions of single excitation current states. A single excitation current state, i.e., a single-spin wave, is defined as:
\begin{equation}
    \ket{\ell} = \dfrac{1}{\sqrt{L}} \sum_j e^{i2\pi\ell j / L}\hat{\sigma}_j^+ \ket{\downarrow,\ldots,\downarrow},
    \label{eq:currstate}
\end{equation}
where $\ell$ is the winding number. We define a superposition of such kind of states as
\begin{equation}
\ket{\Psi_{\Lambda}}=\mathcal{N}\sum_{\ell \in \Lambda} e^{i\phi_{\ell}}\ket{\ell},
\label{eq:superp}
\end{equation}
where $\Lambda$ is the set of winding numbers that participate to the superposition, $\mathcal{N}$ is a normalization constant, and $\phi_{\ell}$ accounts for the possible relative phases between the current states.
The superposition of two current states is
\begin{equation}
    \ket{\Psi^{(2)}} = \dfrac{1}{\sqrt{2}} \big( \ket{\ell_1} + e^{i\phi_{21}}\ket{\ell_2} \big),
    \label{eq:2states_sup}
\end{equation}
where $\phi_{21} \equiv \phi_{\ell_2} - \phi_{\ell_1}$. The population distribution $P_j^{(2)}=|\braket{\Psi^{(2)}|j}|^2$ corresponding to such state is
\begin{equation}
P_j^{(2)} 
= \frac{1}{L} \bigg\{ 1 + \cos \Big[ \dfrac{2\pi (\ell_2 - \ell_1) j}{L} + \phi_{21} \Big] \bigg\} .
\label{eq:pop_ell12}
\end{equation}
For the class of states (\ref{eq:2states_sup}), the maximum value of the on-site magnetization $M_S^z = \braket{ \hat{\sigma}_S^z}_{\psi}=2P_S^{(2)}-1$ is ${4}/{L} - 1$, thus for $L>4$ it is always negative. For the sake of simplicity, we consider $L>4$ and observe that $\mathcal{M}_{zz}=-\Delta M_S^z = \Delta |M_S^z|$, so $\mathcal{M}_{zz} \lessgtr 0$ corresponds to $\Delta \lessgtr 0$. 

\begin{figure*}[!t]
\centering
\includegraphics[width=\textwidth]{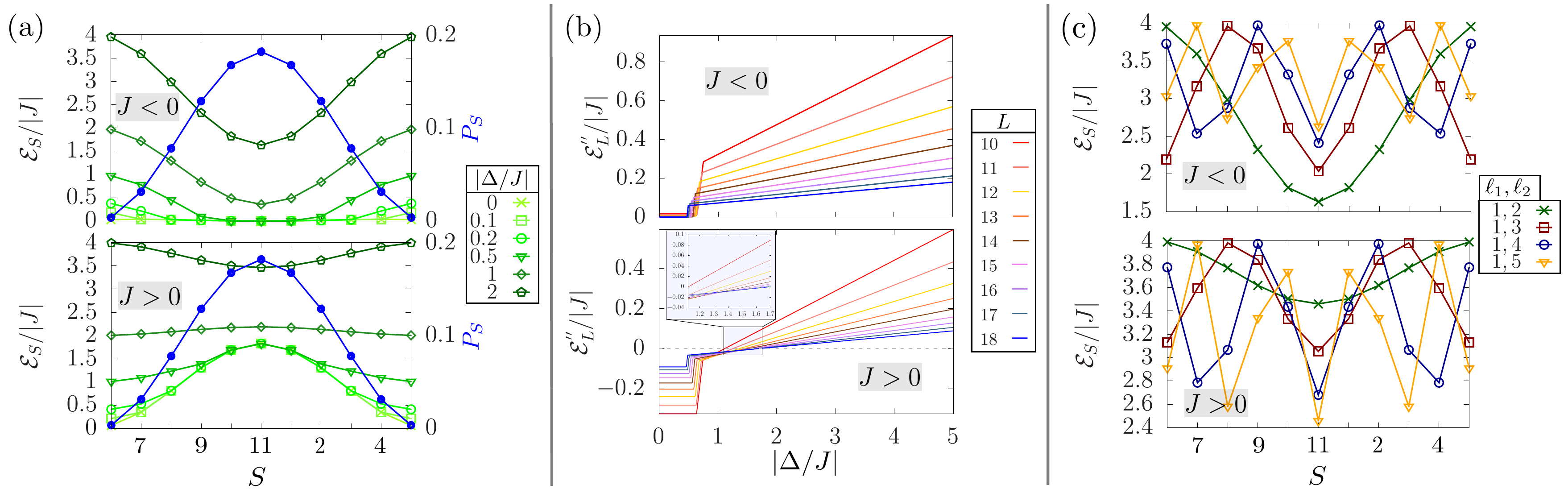}
\caption{Numerical study of the LE distribution for a superposition two-current states with $\phi_{21}=0$, in a system described by the XY Hamiltonian~\eqref{eq:XY_NN_ham} with $\Delta\leq 0$. (a): The LE distribution $\mathcal{E}_S$ (green) compared with the population distribution $P_S$ (blue) for different values of $|\Delta/J|$. (b): The convexity quantifier $\mathcal{E}_L^{''}$ in the site $S=L$, as a function of $|\Delta/J|$, for different system sizes. (c): The LE distribution for $\Delta/|J|=-2$ and different choices of $\ell_1$ and $\ell_2$. In (a) and (b), the superposed current states are $\ell_1=1$ and $\ell_2=2$. In (a) and (c) we fix a system size $L=11$. Top and bottom plots are for $J<0$ and $J>0$, respectively.}
\label{fig:ergdistr}
\end{figure*}

Thus the LE is given by
\begin{equation}
  \dfrac{\mathcal{E}_S}{2 |J|} = \sqrt{(C_S^{xx})^2 \!+ \!(C_S^{xy})^2} + \textrm{sgn}(J) \, C_S^{xx} , \quad \mbox{if } \Delta \!\geq \! 0,
\end{equation}
where $C_S^{xx} = \braket{\hat{\sigma}_S^x \otimes \hat{X}_E}_{\psi}$ and $C_S^{xy} = \braket{\hat{\sigma}_S^x \otimes \hat{Y}_E}_{\psi}$, as defined in Eqs.~\eqref{eq:Corr12}.
For negative $\Delta$, we first observe that the condition $\sqrt{\mathcal{M}_{xx}^2 + \mathcal{M}_{xy}^2} \lessgtr |\mathcal{M}_{zz}|$ can be written in terms of correlation functions as  $|\Delta/J| \lessgtr g_S$, where $g_S = \sqrt{(C_S^{xx})^2 + (C_S^{xy})^2}/|M_S^z|$. Thus we obtain
\begin{equation}
\dfrac{\mathcal{E}_S}{2 |J|}=
\begin{cases}
 \sqrt{(C_S^{xx})^2 \!+\! (C_S^{xy})^2} + \textrm{sgn}(J)C_S^{xx} & \textrm{if} \; \bigg| \dfrac{\Delta}{J} \bigg| \! < \! g_S , \\ \\
 \bigg| \dfrac{\Delta M_S^z}{J} \bigg| + \textrm{sgn}(J)\, C_S^{xx} &  \textrm{if} \; \bigg| \dfrac{\Delta}{J} \bigg| > g_S .
\end{cases}
\label{eq:Jnega}
\end{equation}
Figure~\ref{fig:sketch_ergotropy} sketches the behavior of the LE in the different Hamiltonian parameters ranges. The explicit calculations of the correlation functions and the on-site magnetization are reported in Appendix~\ref{app:corr_func_sup}.

\subsection{Local ergotropy distribution}

We now study the distribution of the LE in the ring:  Fig.~\ref{fig:ergdistr}, reports our analysis for finite values of $L$ and for superpositions of two current states, as in Eq~\eqref{eq:2states_sup}. Let us first focus on the superposition of $\ell_1=1$ and $\ell_2=2$ with a relative phase  $\phi_{21}=0$. The corresponding population distribution~\eqref{eq:pop_ell12} is found to be peaked in $j=L$, with a maximum value $P_L = 2/L$. For even $L$, its minimum is at $j=L/2$, while for odd $L$, it develops two minima at $j=(L\pm 1)/2$. We consider $L>4$ for which $\mathcal{M}_{zz}\propto \Delta$. For $\Delta > 0$, the LE coincides with those obtained with $\Delta = 0$. For this reason, we consider only $\Delta \leq 0$. 

In Fig.~\ref{fig:ergdistr}(a) we report the behavior of the LE distribution for different values of $\Delta$, comparing the cases $J>0$ and $J<0$. As a reference, we also plot the population distribution in blue. For negative $J$ (top panel) we observe that the ergotropy has an opposite behavior compared to $P_j^{(2)}$: when the population is maximum, the ergotropy is minimum and vice-versa. ${\mathcal E}_S$ is larger as $|\Delta|$ increases. Regarding the positive $J$ case (bottom panel), the behavior of the ${\mathcal E}_S$ is different. For small values of $\Delta$, the ${\mathcal E}_S$ is peaked at $S=L$ and has the minimum at the population minimum. For small values of $|\Delta|$, the ${\mathcal E}_S$ around the peak remains unchanged. Differently from the $J<0$ case, here the shape of the distribution changes as $|\Delta|$ 
noticeably increases, with the peak that moves to the minimum population site. The change in the shape of the LE distribution is related to a competition between local spin exchange and field energy scales (see Appendix~\ref{app:per_site_energy}). When the hopping and the field have opposite sign, the energy distribution of the system has different shapes depending on which energy scale is dominant. The same happens to the LE distribution.

To compare the shape of the distribution for different values of the Hamiltonian parameters, we refer to the LE convexity 
\begin{equation}
    \mathcal{E}_S^{''}=(\mathcal{E}_{S+1} - \mathcal{E}_S) - (\mathcal{E}_S - \mathcal{E}_{S-1}) \, .
    \label{eq:convexity}
\end{equation}
For $\mathcal{E}_S^{''}>0$ the distribution is convex in $S$, while for $\mathcal{E}_S^{''}<0$ it is concave in $S$. Figure~\ref{fig:ergdistr}(b) displays $\mathcal{E}_{S}^{''}$ for $S=L$.
For $J<0$ (top panel), we observe that $\mathcal{E}_L^{''}$ is always positive, for any value of $|\Delta|$ and $L$ considered. When $J>0$ (bottom panel), $\mathcal{E}_L^{''}$ changes sign for increasing values of $|\Delta/J|$, the distribution experiences a change  from concave to convex shape in $S=L$. The values for which the ergotropy distribution changes shape can be estimated as $|\Delta/J|=\cos k_1 + \cos k_2$, with $k_{a}=2\pi\ell_{a}/L$. They correspond to the threshold energy values above and below which the spin exchange or the field energy contribution dominate (see Appendix~\ref{app:per_site_energy}). For both negative and positive $J$, the behavior of $\mathcal{E}_L^{''}$ is characterized by three different regimes: flat, fast, and slow linear growth. The transition between the three mentioned regimes is related to the values of $g_L$ and $g_{L\pm 1}$. Indeed, it can be observed that the flat regime is obtained for $|\Delta/J|<g_{L},g_{L\pm 1}$, the fast growing linear regime for $|\Delta/J|<g_{L}$, $|\Delta/J|>g_{L\pm 1}$, and the slow growing linear regime for $|\Delta/J|>g_{L},g_{L\pm 1}$. 
In the first case, the LE does not depend on $|\Delta|$ [see Eq.~\eqref{eq:Jnega}], thus the convexity~\eqref{eq:convexity} is flat in $|\Delta|$; in the fast linear regime, $\mathcal{E}_{L\pm 1}\propto |\Delta|$ and so the positive contributions to $\mathcal{E}_L^{''}$~\eqref{eq:convexity} are linear in the detuning. In the slow linear regime, both $\mathcal{E}_L, \mathcal{E}_{L\pm 1}\propto |\Delta|$, thus both positive and negative contributions to $\mathcal{E}_L^{''}$~\eqref{eq:convexity} are linear in the detuning. This is the reason why the slope of the latter is reduced with respect to those of the fast linear regime. 

\begin{figure}[!t]
\centering
\includegraphics[width=\linewidth]{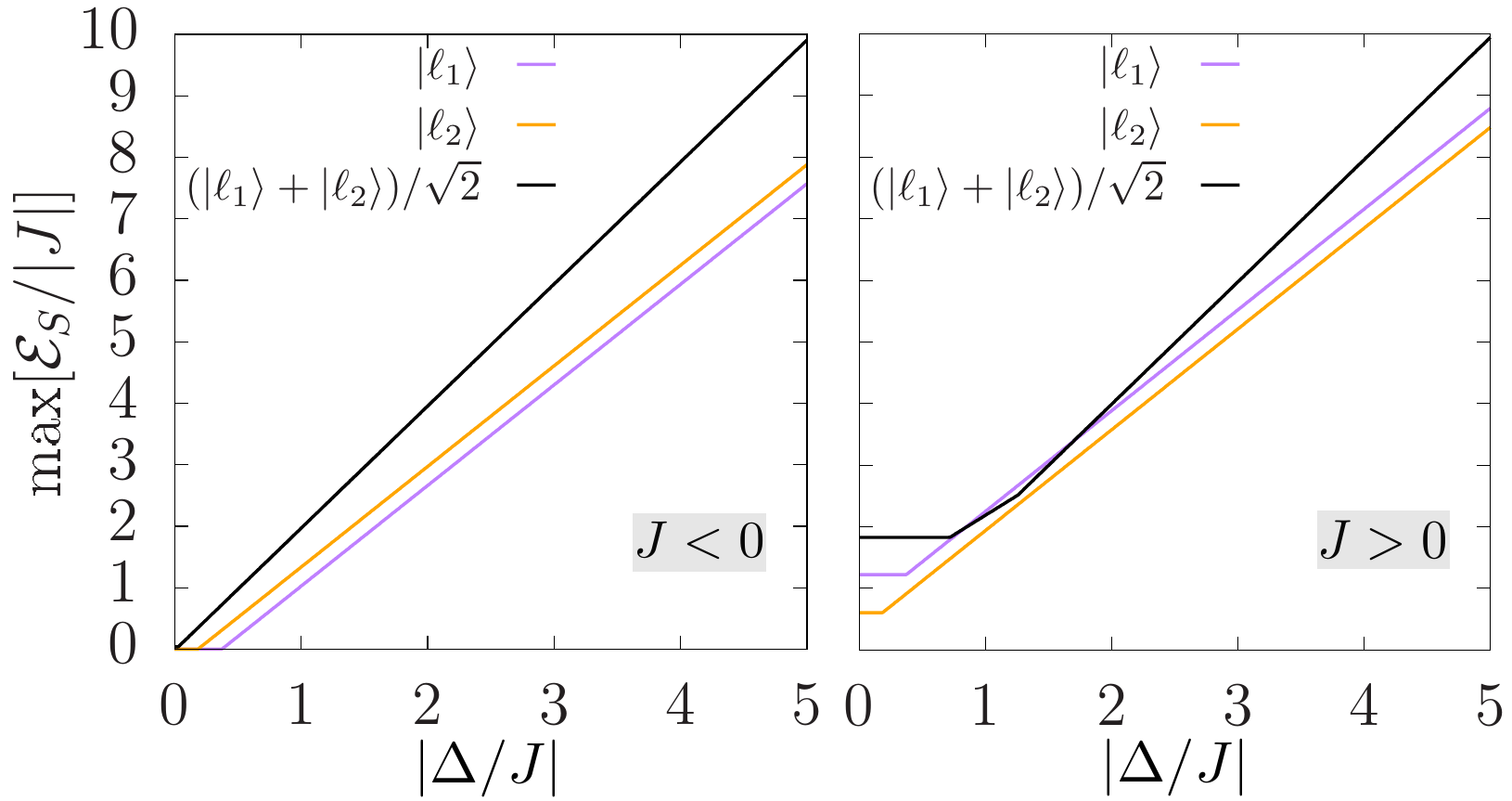}
\caption{Maximum LE over the ring as a function of $|\Delta/J|$, for $\Delta \leq 0$, for the single current-states $\ell_1=1$ and $\ell_2=2$, and for their superposition. In the case of single-current states the LE is homogeneous, its maximum is independent on the site. The number of sites is $L=11$. The two panels are for $J<0$ (left) and $J>0$ (right).}
\label{fig:maximum_ergotropy}
\end{figure}

We also analyze the ergotropy distribution for different superposed current states $\ell_1$ and $\ell_2$, with $\phi_{21}=0$. In particular, we fix $\ell_1$ and we increase $\ell_2$. For odd $L$, the increase of $\ell_2 - \ell_1$ causes a change in the population distribution~\eqref{eq:pop_ell12} with an appearance of local maxima, the global maximum is still present in $S=L$. Figure~\ref{fig:ergdistr}(c) reports the ergotropy distribution for different values of $\ell_2 -\ell_1$, $L=11$ and fixed $\Delta/|J|=-2$. We change $\ell_2$ and keep $\ell_1=1$ fixed. Increasing $\ell_2$, we note that the LE follows the opposite behavior of the population. The positive and negative $J$ cases differ for the depth of the minima. Indeed, for $J<0$ (top panel), the depth of the minima increases as soon as $\ell_2-\ell_1$ decreases; conversely, for $J>0$ (bottom panel), the depth increases for increasing values of $\ell_2 - \ell_1$. 

The extractable energy of superpositions of current states is localized in specific areas of the ring. On the other hand, the energy distribution of a single current state is homogeneous. To study whether and how a superposition of current states can be energetically advantageous with respect to single current states, we compare the maximum LE over the ring of the superposition state with those of the single superposed states, see Fig.~\ref{fig:maximum_ergotropy}. In the asymptotic limits $|\Delta/J| \gg 1$ and $|\Delta/J|\ll 1$, the maximum LE of the superposition state is clearly dominant with respect to those of the superposed states. In particular, for $|\Delta/J|=0$, it is the sum of the LEs of the two superposed states. The LE of the superposition state considered is mainly dominant with respect the two current states considered, although there are values of the Hamiltonian parameters for which the LE of one of the two superposed current states dominates. A detailed analysis of the LE of single current states and its comparison with those of superposition states is reported in Appendix~\ref{app:comparison}.

\begin{figure}[!h]
\centering
\includegraphics[width=\linewidth]{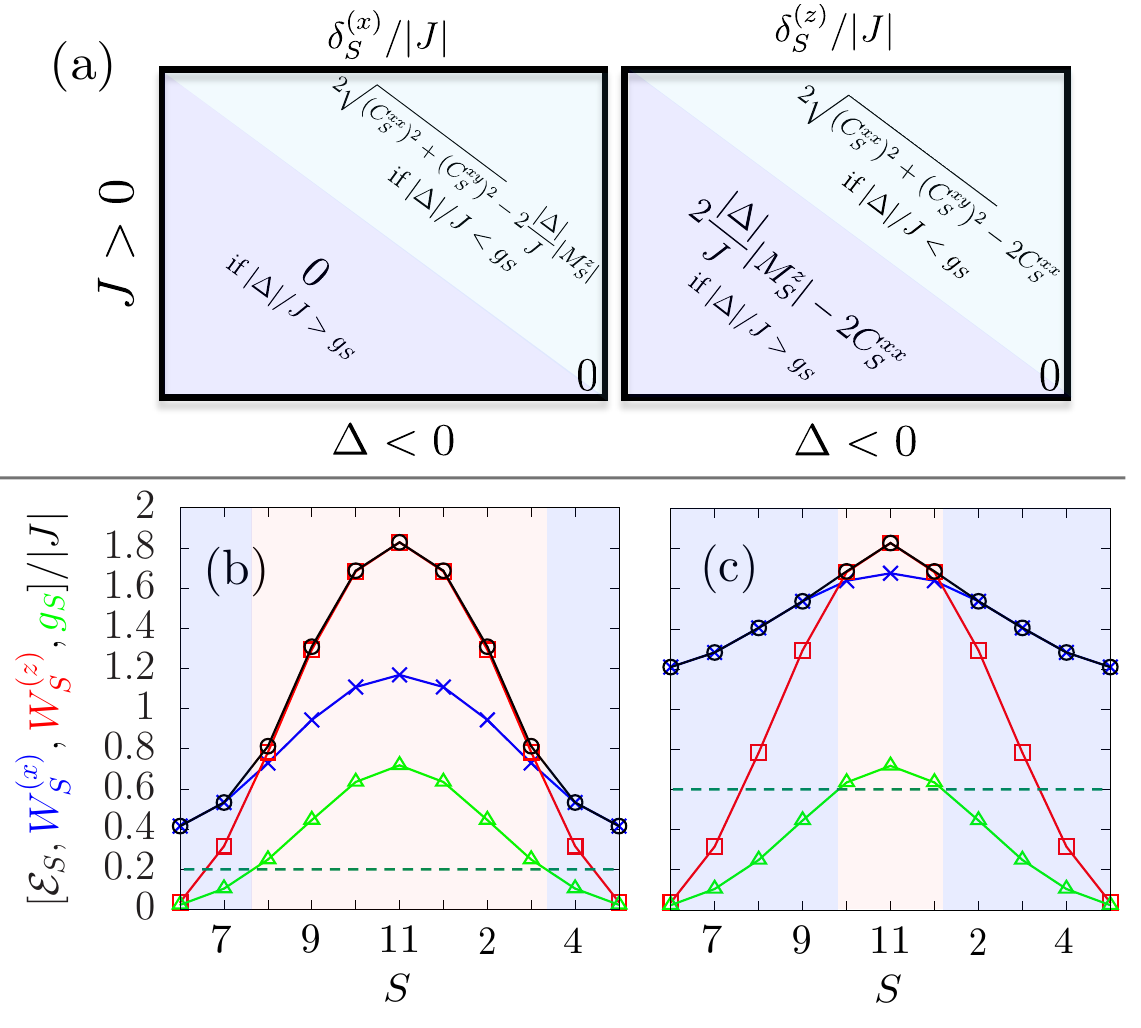}
\caption{A comparison between LE and extractable energies through local rotations, for negative $\Delta$ and positive $J$, for a superposition of two quantum current states with $L>4$. (a): The discrepancy between LE and extractable energies through local transformations [cf.~Eq.~\eqref{eq:distance}] in the $(\Delta<0,J>0)$ plane. (b): The distribution of the LE, the extractable energies through $X$ and $Z$ transformations, and $g_S = \sqrt{(C_S^{xx})^2 + (C_S^{xy})^2}/|M_S^z|$. Here we set $|\Delta|/J=0.2$, choose a superposition of $\ell_1 = 1$ and $\ell_2=2$ states, and fixed $L=11$. (c): Same as (b), but with $|\Delta|/J=0.6$.}
\label{fig:optimal}
\end{figure}

\begin{figure*}[!t]
\centering
\includegraphics[width=0.85\textwidth]{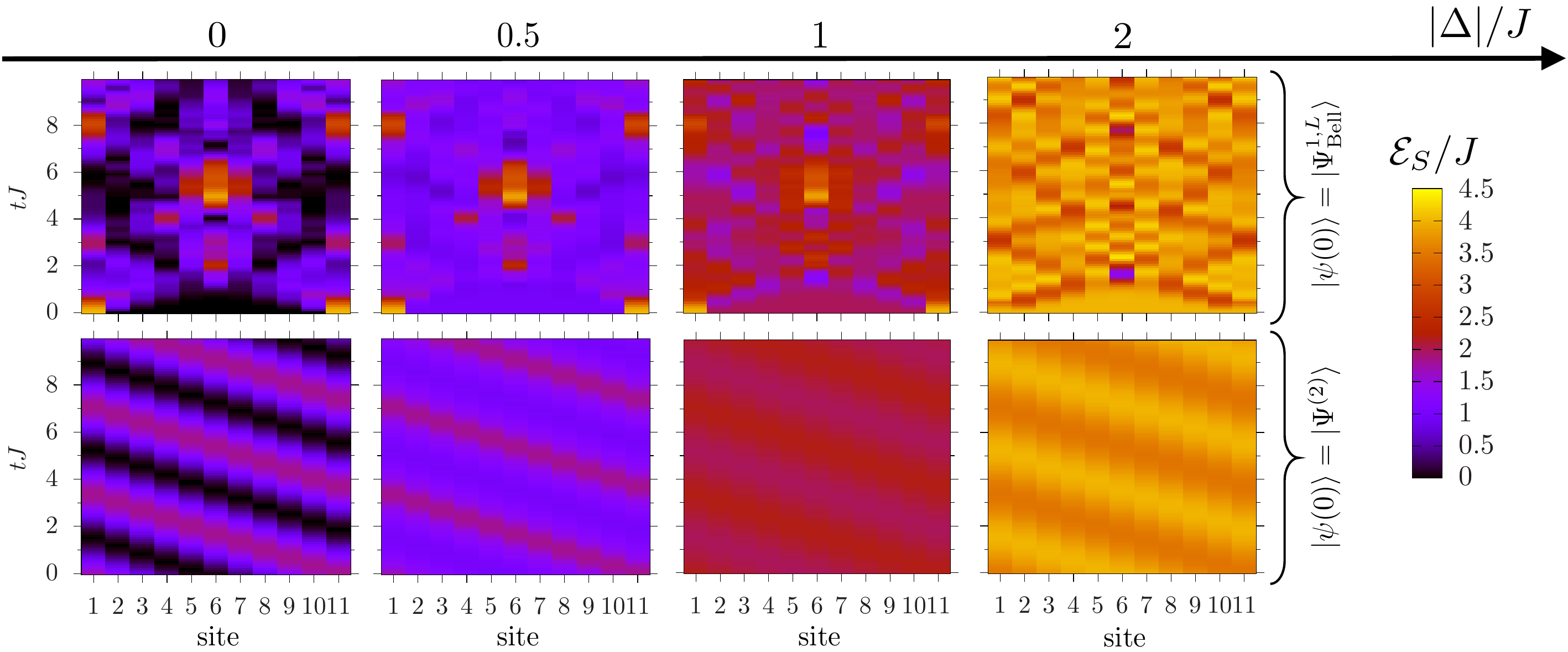}
\caption{Dynamics of the LE for $J>0$ and $\Delta \leq 0$. Top panels show the dynamics for a given initial state $\ket{\psi(0)}=\ket{\Psi_{\rm Bell}^{L,1}}$ [cf.~Eq.~\eqref{eq:Bell_state}] and various values of the ratio $|\Delta|/J$. Bottom panels are the same, but starting for an initial state $\ket{\psi(0)}=\ket{\Psi^{(2)}}$, with $\ell_1=1$, $\ell_2=2$, and $\phi_{21}=0$. The size of the system is $L=11$. 
}
\label{fig:E_vs_j_t}
\end{figure*}

\subsection{Optimal transformations}
\label{sec:opt_transf}

Important insight on the LE is obtained by working out the local optimal transformations $\hat{U}_S$ maximizing $W_S$ [see Eq.~\eqref{eqref:optimal}]. To this purpose, we parametrize them as 
\begin{equation}
    \hat{U}_S=\hat{\mathbb{1}}_S \cos (\theta / 2) - i (\boldsymbol{n}
\cdot \hat{\boldsymbol{\sigma}}_S)\sin(\theta/2).
\end{equation}
The procedure consists in finding the  optimal angle $\theta$ and three components of the unit vector  $\boldsymbol{n}$, for each site and values of the Hamiltonian parameters, with the goal of saturating LE. We consider an heuristic approach and select  $\hat{U}_S=\hat{\sigma}_S^x,\hat{\sigma}_S^z$, that we identify as $X$ and $Z$ transformations. The extractable work through local $X$ and $Z$ transformation is reported in Appendix~\ref{app:XZ_extractable}. Comparing Eq.~\eqref{eq:WsXZ} with Eqs.~\eqref{eq:E_S<>}, we observe that there is a clear relation between them and the LE. In particular, for $C_S^{xy}=0$, the only possible optimal transformations are $X$ and $Z$, for any value of $\Delta$ and $J$. 
We introduce the parameter
\begin{equation}
\delta_S^{(\alpha)}=\mathcal{E}_S - W_S^{(\alpha)},
    \quad
    \alpha=x,z,
    \label{eq:distance}
\end{equation}
where $W_S^{(\alpha)} = E[\ket{\psi}]-E[\hat{\sigma}_S^{\alpha}\ket{\psi}]$. The parameter $\delta_{S}^{(\alpha)} \! \geq \! 0$ and vanishes  when $\hat{\sigma}_S^{\alpha}$ is the optimal transformation.

In Fig.~\ref{fig:optimal}(a) we report $\delta_S^{(\alpha)}$, quantifying the performance of 
$X$ and $Z$ transformations, for negative $\Delta$ and positive $J$. The values of  $\delta_S^{(\alpha)}$ are computed using the LE and the extractable energies reported respectively in Figs.~\ref{fig:sketch_ergotropy} and~\ref{fig:Wzx_scheme}. We observe that the $X$ transformation is optimal for $|\Delta|/J > g_S$, while $Z$ transformation is optimal for $|\Delta|/J \leq g_S$ and states with $C_{S}^{xy}=0$ and positive $C_S^{xx}$. For a superposition of two current states with $\phi_{21}=0$, we can observe that $C_S^{xy}$ is exactly zero for $S=L$ [see Eq.~\eqref{eq_Cxy_two} in Appendix~\ref{app:corr_func_sup}]. Far from $S=L/2$ and for $\cos k_1 + \cos k_2 \gg |\cos k_1 - \cos k_2|$, we have $C_S^{xx}\gg C_S^{xy}$ and the $Z$ transformation can be identified as quasi optimal (it is exactly optimal in $S=L$). This is the case of a superposition $\ell_1=1$ and $\ell_2=2$. 

Figure~\ref{fig:optimal}(b) shows the ergotropy distribution together with the extractable energies and $g_S$ for a superposition of two current states and $|\Delta|/J=0.2$. For $|\Delta|/J>g_S$, $X$ is the optimal transformation. The sites in which this condition is satisfied are far from $S=L$. Conversely, for $|\Delta|/J<g_S$, the $X$ transformation is not anymore optimal, while the discrepancy between $\mathcal{E}_S$ and $W_S^{(z)}$ is minimal and $Z$ can be considered quasi optimal. The latter is exactly optimal in $S=L$, where $C_L^{xy}=0$. In Fig.~\ref{fig:optimal}(c), we show the same type of plot for $|\Delta|/J=0.6$; since $|\Delta|/J$ is increased, the number of sites for which $g_S>|\Delta|/J$ is reduced. The optimal transformation is $X$ everywhere except in $S=L$ and its nearest neighbors. Finally, we mention that for values of $|\Delta|/J>g_S$ on all the sites, the optimal transformation is $X$ everywhere.

We observe that the threshold for which the $X$ transformation becomes optimal is $|\Delta \, M_S^z|= |J|\sqrt{(C_S^{xx})^2 + (C_S^{xy})^2}$, i.e. when local field and spin exchange related energy scales are comparable.
Specifically, the $X$ transformation is optimal when the field energy scale is dominant, while $Z$ is quasi optimal in the opposite case. The competition between field related and correlations related energy scales is crucial to understand which is the optimal transformation.

\section{Chiral flows of ergotropy}
\label{sec:chiral_ergo}

In this section, we study the dynamics of the LE, by initializing the system in a superposition of two current states, $\ket{\psi(0)}=\ket{\Psi^{(2)}}$, and evolving it through the Hamiltonian~\eqref{eq:XY_NN_ham}. We use $\ell_1=1$ and $\ell_2=2$, and set the relative phase $\phi_{21}$ of the initial state to zero. Since the two superposed states are Hamiltonian  eigenstates~\cite{perciavalle2024quantum}, a relative phase between the two current states is set. The LE dynamics of the latter state is compared with those of another state in which two sites are prepared in a Bell state and the rest of the ring is in the $\ket{\downarrow}$ state. We consider the Bell state to be in the sites $1$ and $L$ and thus the many-body state reads 
\begin{equation}
\ket{\Psi_{\rm Bell}^{1,L}} = \frac{1}{\sqrt{2}}\Big( \ket{\uparrow_1,\downarrow_2,\ldots,\downarrow_L} + \ket{\downarrow_1,\downarrow_2,\ldots,\uparrow_L} \Big),
\label{eq:Bell_state}
\end{equation}
which is of interest for the study of entanglement propagation in quantum systems~\cite{amico2004dynamics}. The energy of such state is $E\bigl[\ket{\Psi_{\rm Bell}^{1,L}}\bigr]=2J + \Delta (2-L)$. We observe that $E\bigl[\hat{\sigma}_j^z\ket{\Psi_{\rm Bell}^{1,L}}\bigr]=-2J + \Delta (2 - L)$ and $E\bigl[\hat{\sigma}_k^x\ket{\Psi_{\rm Bell}^{1,L}}\bigr]=2J + \Delta (4-L)$ where $j=1,L$ and $k\neq 1,L$.

Figure~\ref{fig:E_vs_j_t} shows the dynamics of the ergotropy for the two different initial states; negative $\Delta$ and positive $J$ are considered. At $t=0$, we observe that the LE of the state $\ket{\Psi_{\rm Bell}^{1,L}}$ is $4J$ in correspondence of the entangled sites $j=1,L$  and it is $2|\Delta|$ in the sites $k\neq 1,L$. Thus, the $\hat{\sigma}_j^z$ rotation is  optimal in correspondence of the entangled sites $j=1,L$, while $\hat{\sigma}_k^x$ is optimal elsewhere. Similarly to what happens to the superposition of two current states, the ring is divided in two areas in which $X$ and $Z$ transformations are good choices. The main difference between the two cases can be observed in the dynamics. Regarding the Bell state, we observe that its nature is broken and the LE spreads along the ring showing a non-trivial dynamics. 
On the other hand, the dynamics of the superposition of current states is characterized by a perfect chiral flow of the LE. 
Different values of negative $\Delta$ affect the LE distribution, thus its dynamics is different. The dynamics is perfectly directional, in absence of dissipation it persists at any time. Increasing the value of $|\Delta|$, the peak of the distribution moves from $S=L$ to $S=(L\pm 1)/2$. Thus, keeping unchanged the state, it is possible to control the position of the peak of the extractable energy and its dynamics by properly adjusting the Hamiltonian parameters. Finally, we mention that the optimal transformation $\hat{U}_S$ depends on the ergotropy distribution. For a superposition state, the distribution is shifted in time but its shape remains unchanged. Thus, the optimal or quasi-optimal transformations are the same as those discussed in Sec.~\ref{sec:opt_transf}; the sites on which they have to be performed to optimize the energy extraction change in time.

\section{Local ergotropy versus long-range spin interactions}
\label{sec:ergo_dipolarXY}
In this section, we study how long-range spin interactions may affect the distribution and the dynamics of the LE. Specifically, we consider a generalized version of the Hamiltonian~(\ref{eq:XY_NN_ham}), which reads
\begin{equation}
\hat{H}=\sum_{i\neq j} \frac{J_{ij}}{2}\left(\hat{\sigma}_i^x \hat{\sigma}_j^x + \hat{\sigma}_i^y \hat{\sigma}_j^y \right) + \Delta \sum_j \hat{\sigma}_j^z,
 \label{eq:Ham_dipolar}
\end{equation} 
where $J_{ij} = 2g / d_{ij}^{\alpha}$ is the hopping strength and $d_{ij}$ is the distance between the sites. Considering a ring-shaped geometry composed of $L$ sites, the distance reads $d_{ij}=2R\sin(\pi |i-j|/L)$, $R$ being the radius of the ring. The case $\alpha=3$ in Eq.~\eqref{eq:Ham_dipolar} can be realized with Rydberg atoms trapped in optical tweezers, in which $\ket{\uparrow}_j$ and $\ket{\downarrow}_j$ are two Rydberg states of opposite parity~\cite{browaeys2020many, barredo2015coherent, morgado2021quantum}. The external field term can be realized by globally coupling the atoms with a microwave field with detuning $\Delta$~\cite{browaeys2020many}. Since Rydberg atoms trapped in optical tweezers can be arranged in various geometries~\cite{bernien2017probing, bluvstein2023logical, schymik2020enhanced, barredo2016atom, pause2024supercharged, pichard2024rearrangement}, the realization of a ring-shaped circuit is experimentally feasible. The nearest-neighbor case is recovered for $\alpha=\infty$. The realization of quantum superpositions of current states has been theoretically proposed using quantum optimal control~\cite{perciavalle2024quantum,koch2022quantum, Khaneja2005optimal, goerz2022quantum} and machine learning techniques~\cite{haug2021machine}: initializing the system in a state with a single fully localized excitation, the detuning is locally modulated and used as control parameter, the desired state is obtained at a certain target time with high fidelity~\cite{perciavalle2024quantum}.

In the case of a generic spin exchange $J_{ij}$, the matrix elements in the one excitation sector are
\begin{align}
& \mathcal{M}_{xx}=-\!\sum_{j \neq S} J_{Sj} \braket{\hat{\sigma}_S^x\hat{\sigma}_{j}^x}_{\psi}, \; \mathcal{M}_{yy}=-\!\sum_{j \neq S} J_{Sj} \braket{\hat{\sigma}_S^y\hat{\sigma}_{j}^y}_{\psi},
\nonumber \\
& \mathcal{M}_{xy}=-\!\sum_{j \neq S} J_{Sj} \braket{\hat{\sigma}_S^x\hat{\sigma}_{j}^y}_{\psi},
\;
\mathcal{M}_{yx}=\!\sum_{j \neq S} J_{Sj} \braket{\hat{\sigma}_S^y\hat{\sigma}_{j}^x}_{\psi},    \nonumber  \\
& \mathcal{M}_{zz}=-\Delta \braket{\hat{\sigma}_S^z}_{\psi},   
\end{align}
(see Appendix~\ref{sec:M_XXmodel} for the complete derivation).  
The LE structure is unchanged and follows Eqs.~\eqref{eq:loc_erg}. We mention that multi-basis two-body correlation functions are experimentally accessible in Rydberg-atom platforms~\cite{bornet2024enhancing}, thus the LE can be measured.

\begin{figure}[!h]
\centering
\includegraphics[width=\linewidth]{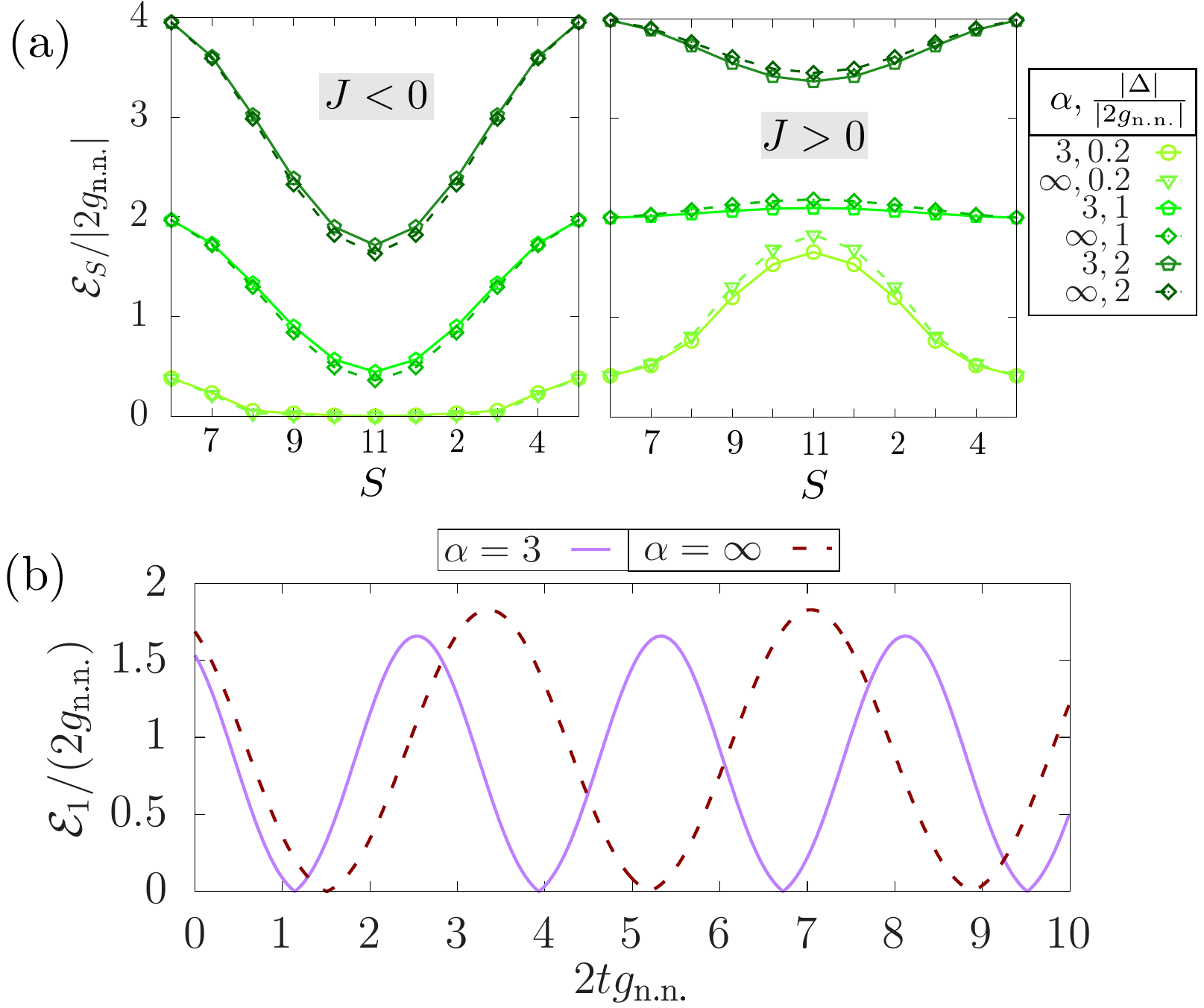}
\caption{The LE of a quantum superposition of current states, for a system with spin exchange interactions. This case is particularly relevant for Rydberg-atom platform. (a): The LE distribution for negative (left) and positive (right) values of $J = 2g_{\rm n.n.}$. Cases with different $|\Delta/2g_{\rm n.n.}|$ and $\alpha=3,\infty$ are compared. (b): The dynamics of the LE on the site $S=1$ for $J>0$ and the two different values of $\alpha=3,\infty$, the detuning $\Delta$ is set to zero. In all panels we choose the initial state as $(\ket{\ell_1=1} + \ket{\ell_2=2})/\sqrt{2}$ and fix $L=11$.}
\label{fig:LRplots}
\end{figure}

In Figure~\ref{fig:LRplots} we compare the LE of a superposition of two current states of an XY model with dipolar spin exchange ($\alpha=3$) and with nearest-neighbor spin exchange ($\alpha=\infty$). Figure~\ref{fig:LRplots}(a) reports the distribution for positive and negative $J=2g_{\rm n.n.}$, where $g_{\rm n.n.} = g/(2R\sin(\pi/L))^{\alpha}$ is the nearest-neighbor spin exchange strength. We see that the presence of dipolar spin exchange does not qualitatively change the ergotropy distribution, while it is slightly shifted in proximity of the site $L$. The main difference between the two cases is observed in the dynamics. A superposition of $\ell_1=1$ and $\ell_2 = 2$ states acquires a relative phase factor $\phi_{21}=(E_{\ell_1} - E_{\ell_2})t$, from which it is possible to extract the velocity at which the information travels along the ring~\cite{perciavalle2024quantum}. Given the Hamiltonian~\eqref{eq:Ham_dipolar} with $\Delta=0$, the energy of the state $\ket{\ell}$ is
\begin{equation}
E_{\ell}=4\sum_{k}\phantom{}^{'}  g_k \cos\left(\dfrac{2\pi\ell k}{L} \right),
\end{equation}
where $\sum_k \!\! \phantom{}^{'} \coloneq  \sum_{k=-\floor{L/2},(k\neq 0)}^{\floor{L/2}}$ for even $L$ and $g_k = g / \big[ 2 R \sin(\pi|k|/L) \big]^{\alpha}$~\cite{perciavalle2024quantum}. Thus, the value of $\alpha$ affects the spectrum of the system and consequently the velocity of the information. 
We report the dynamics of a specific superposition state in Fig.~\ref{fig:LRplots}(b), where we observe a bigger oscillation frequency of the LE for $\alpha=3$ with respect to the nearest-neighbor case, thus meaning a faster transfer of extractable energy along the ring.

\section{Outlook and conclusions}
\label{sec:outlook}

We studied the energetics of spin-wave states, with a specific focus on superposition states. An important goal for our work is to theoretically demonstrate the controlled manipulation of the extractable energy in the system. To this end, we refer to the local ergotropy (LE) as figure of merit for the extractable energy~\cite{salvia2023optimal,mukherjee2016presence,imai2023work, castellano2024extended}, which can be computed using a closed formula if the local transformations act on a single qubit~\cite{salvia2023optimal}. As for the physical system, we consider a prototypical quantum spin-$1/2$ XY Hamiltonian model in an external field.

We show that, when using a superposition of current states, by tuning the system's parameters $J$ and $\Delta$ in Eq.~\eqref{eq:XY_NN_ham}, we can indeed control the ergotropy locally. 
The LE is inhomogeneous through the system. In particular, a single maximum appears for a superposition of two states, overcoming the ergotropy of current states consisting of single winding number (see Fig.~\ref{fig:maximum_ergotropy}). The transformations that permit to extract the maximum amount of energy can be identified. Surprisingly, for complex states such as the superposition of two currents, we find that simple Pauli rotations are optimal or quasi-optimal (see Fig.~\ref{fig:optimal}). 
Our states are engineered by superposing the eigenstates of the Hamiltonian. The study of the dynamics demonstrates that the ergotropy of superpositions of two current states have a fully controlled dynamics, characterized by persistent flows of ergotropy (see Fig.~\ref{fig:E_vs_j_t}). 

Rydberg atoms provide an interesting potential platform to investigate our predictions. In this case, spin states are two internal Rydberg states with opposite parity, $J$ can be controlled through the interatomic distance, and $\Delta$ can be controlled through the detuning of a microwave field that globally couples the atoms. To make contact with the Rydberg physics, we studied the impact of long-range interactions on the LE distribution. We found that, unlike its spatial distribution, the ergotropy dynamics is characterized by a velocity that depends on dipolar interactions.
We remark that, in Rydberg-atom platforms, multi-basis and multi-body correlation functions have been experimentally measured~\cite{bornet2024enhancing}, thus the LE and, more in general, the extractable energy could be in principle accessible.
Ultracold atoms, superconducting qubits, and trapped ions provide further options for physical implementations~\cite{roushan2016chiral,chiaro2022direct,maier2019environment,blatt2012quantum, polo2024perspective, bloch2008many}.

It would be tempting to engineer suitable integrated circuits able to extract and store the energy distribution of the system. 
As a follow up of the present study, we think that including disorder, generalizing to different classes of physical systems, and exploring the extractable energy from higher dimensional subsystems would provide interesting lines for future investigations. 

\acknowledgements

We thank Oliver Morsch, Enrico C. Domanti, Giampiero Marchegiani, and Frederico Brito for useful discussions.  The Julian Schwinger Foundation grant JSF-18-12-0011 is acknowledged.

\appendix

\section{Derivation of the $\mathcal{M}$ matrix elements for the nearest-neighbor model}
\label{app:M_XYmodelNN}

We focus on the model described in Eqs.~\eqref{eq:XYmodel} and start with the computation of the $\mathcal{M}_{xx}$ matrix element~\cite{salvia2023optimal} on a generic state $\ket{\psi}$:
\begin{equation}
    \mathcal{M}_{xx}=-\left(r_x h_x + \tfrac{1}{2}\textrm{Tr}_E\{\rho_E^{(x)}\hat{V}_E^{(x)}\}\right),
\end{equation}
with $r_x h_x = \tfrac12 \Tr_S\{\hat{\sigma}_S^{x}\rho_S\}\Tr_S\{\hat{\sigma}_S^x \hat{H}_S\}$, $\rho_E^{(x)}=\Tr_S\{\hat{\sigma}_S^x\rho_{SE}\}$ and $\hat{V}_{E}^{(x)}=\Tr_S\{\hat{\sigma}_S^{x}\hat{V}_{SE}\}$, cf.~Eqs.~\eqref{eq:Mmatr_def}-\eqref{eq:Mmatr_def2}. 
Using the known relation for the Pauli matrices
\begin{equation}
\hat{\sigma}^\alpha_{k}\hat{\sigma}_l^\beta = i\epsilon^{\alpha \beta \gamma}\delta_{k,l}\hat{\sigma}_k^\gamma, \qquad
(k,l=1,\ldots,L),
\label{eq:sigma-comm}
\end{equation}
we observe that $\textrm{Tr}_S\{\hat{\sigma}_S^{x}\hat{H}_S\}=-2i\Delta\textrm{Tr}_S\{\hat{\sigma}_S^{y}\}=0$ and thus we can rewrite the $\mathcal{M}_{xx}$ matrix element as
\begin{equation}
\mathcal{M}_{xx}=-\tfrac{1}{2}\textrm{Tr}_E \left[ \Tr_S\{\hat{\sigma}_S^x\rho_{SE}\}  \Tr_S\{\hat{\sigma}_S^{x}\hat{V}_{SE}\}\right].
\label{eq:Mxxcomp}
\end{equation}
We compute
\begin{eqnarray}
\Tr_S\{\hat{\sigma}_S^{x}\hat{V}_{SE}\} & = & J \Tr_S [\hat{\mathbb{1}}_S (\hat{\sigma}_{S-1}^x + \hat{\sigma}_{S+1}^x)\nonumber\\ && + (\hat{\sigma}_S^x\hat{\sigma}_S^y)(\hat{\sigma}_{S+1}^y + \hat{\sigma}_{S-1}^y)] \nonumber \\& = & 2 J (\hat{\sigma}_{S-1}^x + \hat{\sigma}_{S+1}^x) \equiv 2 J \hat{X}_E,
\end{eqnarray}
where we introduced $\hat{X}_E = \hat{\sigma}_{S-1}^x + \hat{\sigma}_{S+1}^x$, $\hat{Y}_E = \hat{\sigma}_{S-1}^y + \hat{\sigma}_{S+1}^y$ and used $\Tr_S \{\hat{\sigma}_S^x\hat{\sigma}_S^y\} = 2i \Tr_S\{ \hat{\sigma}_S^z\}=0$. Coming back to Eq.~\eqref{eq:Mxxcomp},
we can finally rewrite it as
\begin{equation}
\mathcal{M}_{xx}=-J\Tr \big\{ \hat{X}_E\hat{\sigma}_S^x\rho_{SE} \big\} = -J\braket{\hat{\sigma}_S^x \hat{X}_E}_{\psi}.  
\end{equation}
The same type of reasoning can be applied to $\mathcal{M}_{xy}$, $\mathcal{M}_{yx}$, $\mathcal{M}_{yy}$, $\mathcal{M}_{zx}$, and $\mathcal{M}_{zy}$, 
obtaining
\begin{equation}
    \mathcal{M}_{jx}=-J\braket{\hat{\sigma}_S^j \hat{X}_E}_{\psi}, \qquad   \mathcal{M}_{jy}=-J\braket{\hat{\sigma}_S^j \hat{Y}_E}_{\psi}, 
   \label{eq:Mdef}
\end{equation}
with $j = \{x,y,z\}$.

Coming to the $\mathcal{M}_{zz}$ matrix element, we can write it as
\begin{eqnarray}
\mathcal{M}_{zz} & = & -\tfrac{1}{2}\left( \Tr_S\{\hat{\sigma}_S^{z}\rho_S\}\Tr_S\{\hat{\sigma}_S^z \hat{H}_S\} \right. \nonumber \\ & & \left. + \textrm{Tr}_E \big[ \Tr_S\{\hat{\sigma}_S^z\rho_{SE}\}  \Tr_S\{\hat{\sigma}_S^z\hat{V}_{SE}\} \big] \right).
\end{eqnarray}
Since $\hat{H}_S=\Delta\hat{\sigma}^z_S$, we have $\Tr_S\{\hat{\sigma}_S^z \hat{H}_S\} = 2\Delta$ and thus 
\begin{equation}
\Tr_S\{\hat{\sigma}_S^{z}\rho_S\}\Tr_S\{\hat{\sigma}_S^z \hat{H}_S\}=2\Delta\braket{\hat{\sigma}^z_S}_{\psi}.    
\end{equation}
On the other hand, using again the relations~\eqref{eq:sigma-comm}, we find $\textrm{Tr}_S \{\hat{\sigma}_S^{z} \hat{V}_{SE} \} = 0$,
which brings us to the expression 
\begin{equation}
    \mathcal{M}_{zz}=-\Delta\braket{\hat{\sigma}^z_S}_{\psi}.
    \label{eq:Mzz_nn}
\end{equation}
Similarly to this, one can also find $\mathcal{M}_{xz}=-\Delta\braket{\hat{\sigma}_S^x}_{\psi}$ and $\mathcal{M}_{yz}=-\Delta\braket{\hat{\sigma}_S^y}_{\psi}$.

We now consider states belonging to a specific excitation sector as $\ket{\psi_{N\textrm{s}}}$ ($N\in \mathbb{N}$ labels the corresponding $N$-excitations sector). The operators $\hat{\sigma}_S^z \hat{X}_E$, $\hat{\sigma}_S^z \hat{Y}_E$, $\hat{\sigma}_S^x$, and $\hat{\sigma}_S^y$ do not conserve the number of excitations. Applying them on $\ket{\psi_{N\textrm{s}}}$ produces a state with a different number of excitations, which is orthogonal to $\ket{\psi_{N\textrm{s}}}$. For this reason, the matrix elements $\mathcal{M}_{zx}$, $\mathcal{M}_{zy}$, $\mathcal{M}_{xz}$ and $\mathcal{M}_{yz}$ are zero. Conversely, the other matrix elements $\mathcal{M}_{xx}$, $\mathcal{M}_{xy}$, $\mathcal{M}_{yx}$, $\mathcal{M}_{yy}$, and $\mathcal{M}_{zz}$ are in principle non zero. In the general case in which the state $\ket{\psi}$ is a superposition of states belonging to different excitation sectors, the vanishing of some of the matrix elements of $\mathcal{M}$ is not guaranteed and one should take into account the complete matrix $\mathcal{M}$ for the computation of the local ergotropy. In this work, we always deal with states belonging to a specific excitation sector $\ket{\psi_{N\textrm{s}}}$. In this case, it is useful to introduce the dimensionless versions of the matrix $\mathcal{M}$ elements, which are given by the correlation functions
\begin{subequations}
\label{eq:Corr12}
\begin{eqnarray}
    C_S^{xx}= & \braket{\hat{\sigma}_S^x \otimes \hat{X}_E}_{\psi_{N\textrm{s}}}, \quad C_S^{yy}= & \braket{\hat{\sigma}_S^y \otimes \hat{Y}_E}_{\psi_{N\textrm{s}}}, \qquad \label{eq:Corr1} \\ C_S^{xy}= & \braket{\hat{\sigma}_S^x \otimes \hat{Y}_E}_{\psi_{N\textrm{s}}}, \quad C_S^{yx}= & \braket{\hat{\sigma}_S^y \otimes \hat{X}_E}_{\psi_{N\textrm{s}}} , \qquad
    \label{eq:Corr2}
\end{eqnarray}
and the on-site magnetization 
\begin{equation}
M_S^z=\braket{\hat{\sigma}_S^z}_{\psi_{N\textrm{s}}}.
\label{eq:magnetZ}
\end{equation}
\end{subequations}
The resulting nonzero matrix elements of $\mathcal{M}$ are
\begin{subequations}
\begin{align}
&\mathcal{M}_{jk} = - JC_S^{jk}, \quad j,k = \{x,y\}, \\
&\mathcal{M}_{zz}=-\Delta M_S^z.
\end{align}
\end{subequations}

Finally, we further specialize to generic states in the one-excitation sector $\ket{\psi_{1s}}$, which can be written as in Eq.~\eqref{eq:psi_1s}. The $xx$ correlation function can be computed using the relation $\braket{\psi_{1\rm{s}}|\hat{\sigma}_i^x  \hat{\sigma}_{j}^x|\psi_{1\rm{s}}} = \braket{\psi_{1\rm{s}}|\hat{\sigma}_i^+ \hat{\sigma}_{j}^- + \hat{\sigma}_i^-  \hat{\sigma}_{j}^+|\psi_{1\rm{s}}}$. The computation brings to
\begin{align}
C_S^{xx} & =  f_S^* f_{S-1} + f_{S-1}^* f_{S} + f_{S}^* f_{S+1} + f_{S+1}^* f_{S} \nonumber\\ & = f_S (f_{S-1}^* + f_{S+1}^*) + f_S^*(f_{S-1} + f_{S+1}).
\label{eq:CSXX}
\end{align}
Moreover, $\braket{\psi_{1\rm{s}}|\hat{\sigma}_i^y  \hat{\sigma}_{j}^y|\psi_{1\rm{s}}} = \braket{\psi_{1\rm{s}}|\hat{\sigma}_i^+ \hat{\sigma}_{j}^- + \hat{\sigma}_i^- \hat{\sigma}_{j}^+|\psi_{1\rm{s}}}$ holds, and thus $C_S^{yy} = C_S^{xx}$.
For the computation of the $xy$ and $yx$ correlations, we use $\braket{\psi_{1\rm{s}}|\hat{\sigma}_i^x  \hat{\sigma}_{j}^y|\psi_{1\rm{s}}} = i\braket{\psi_{1\rm{s}}|\hat{\sigma}_i^+ \hat{\sigma}_{j}^- - \hat{\sigma}_i^-\hat{\sigma}_{j}^+|\psi_{1\rm{s}}}$.
and we obtain
\begin{equation}
C_S^{xy} = i ( f_S^* (f_{S-1}+f_{S+1}) - f_S (f_{S-1}^*+f_{S+1}^*)),
\label{eq:CS_XY}
\end{equation}
while $C_S^{yx} = -C_S^{yx}$, since $\braket{\psi_{1\rm{s}}|\hat{\sigma}_i^x  \hat{\sigma}_{j}^y|\psi_{1\rm{s}}} = - \braket{\psi_{1\rm{s}}|\hat{\sigma}_i^y  \hat{\sigma}_{j}^x|\psi_{1\rm{s}}}$.
Concerning the on site magnetization, we compute it directly:
\begin{equation}
M_S^z = \sum_{j,k} f_j^* f_k \braket{j|\hat{\sigma}_S^z|k} = |f_S|^2 - \sum_{j\neq S} |f_j|^2=2|f_S|^2 - 1, 
\label{eq:MSZ}
\end{equation}
where we used $\hat{\sigma}^z\ket{\uparrow}=+\ket{\uparrow}$ and $\hat{\sigma}^z\ket{\downarrow}=-\ket{\downarrow}$.
These results bring us to the following relations between the nonzero matrix elements of $\mathcal{M}$, when evaluated on $\ket{\psi_{\rm 1s}}$:
$\mathcal{M}_{xx} = \mathcal{M}_{yy}$ and $\mathcal{M}_{xy} = -\mathcal{M}_{yx}$, while $\mathcal{M}_{zz} = - \Delta M^z_S$.

\section{Derivation of the local ergotropy for quantum superposition of current states}
\label{app:corr_func_sup}

We consider the generic superposition of current states belonging to the set $\Lambda$ in Eq.~\eqref{eq:superp}, namely
\begin{equation}
\ket{\Psi_{\Lambda}} = \sum_j \Big (\dfrac{\mathcal{N}}{\sqrt{L}}\sum_{\ell \in \Lambda}e^{i(2\pi \ell j / L+\phi_{\ell})} \Big) \ket{j}.
\end{equation}
We first compute the correlation functions using the formulas found in the previous appendix, obtaining
\begin{subequations}
\label{eq:C-M-corr}
\begin{eqnarray}
C_S^{xx} & \! = \! & \dfrac{4\mathcal{N}^2}{L} \!\! \sum_{\ell,\ell ' \in \Lambda} \! \cos k_{\ell '} \cos \! \big[ (k_{\ell} \!-\! k_{\ell '})S \! + \! (\phi_{\ell} \!-\! \phi_{\ell '}) \big], \qquad \\
C_S^{xy} & \! = \! & \dfrac{4\mathcal{N}^2}{L} \! \! \sum_{\ell,\ell ' \in \Lambda} \! \cos k_{\ell '} \sin \! \big[(k_{\ell} \!-\! k_{\ell '})S \! + \! (\phi_{\ell} \!-\! \phi_{\ell '}) \big], \qquad
\label{eq:C_S^xy} \\
M_S^z & = & \dfrac{2\mathcal{N}^2}{L}\bigg\lvert\sum_{\ell \in \Lambda}e^{i(k_{\ell} S + \phi_{\ell})}\bigg\rvert^2  - 1,
\end{eqnarray}
\end{subequations}
where $k_{\ell} = 2\pi\ell/L$. Once we know the correlation functions, we can compute $\mathcal{M}$ matrix elements. We consider the nearest-neighbor model~\eqref{eq:XY_NN_ham}, thus $\mathcal{M}_{xx}=-JC_S^{xx}$, $\mathcal{M}_{xy}=-JC_S^{xy}$ and $\mathcal{M}_{zz}=-\Delta M_S^{z}$. The two possible values of the LE  defined in Eqs.~\eqref{eq:loc_erg} are:
\begin{subequations}
\label{eq:E_S<>}
\begin{eqnarray}
\mathcal{E}_S^{>} & \equiv & 2 \Big[ \sqrt{\mathcal{M}^2_{xx} + \mathcal{M}^2_{xy}} - \mathcal{M}_{xx} \Big] \nonumber \\
& = & 2 \Big[ J \sqrt{(C_S^{xx})^2 + (C_S^{xy})^2} + J C_S^{xx} \Big],\\  
\mathcal{E}_S^{<} & \equiv & 2 \big( \vert \mathcal{M}_{zz} \vert - \mathcal{M}_{xx} \big)
= 2 \big( |\Delta \, M_S^z| + J C_S^{xx} \big), \quad
\end{eqnarray}
\end{subequations}
in such a way that $\mathcal{E}_S(\mathcal{M}_{zz}) \geq 0 = \mathcal{E}_S^{>}$ and 
\begin{equation}
    \mathcal{E}_S(\mathcal{M}_{zz} \! < \! 0) = \begin{cases}
    \mathcal{E}_S^{>} & \mbox{if } \sqrt{\mathcal{M}_{xx}^2 \!+\! \mathcal{M}_{xy}^2} > |\mathcal{M}_{zz}|, \\
    \mathcal{E}_S^{<} & \mbox{if } \sqrt{\mathcal{M}_{xx}^2 \!+\! \mathcal{M}_{xy}^2} < |\mathcal{M}_{zz}|.
    \end{cases}
\end{equation}
Substituting Eqs.~\eqref{eq:C-M-corr} in Eqs.~\eqref{eq:E_S<>}, we get
\begin{align}
&\mathcal{E}_S^{>} = \dfrac{8 J \mathcal{N}^2}{L} \Bigg\{ \sum_{\ell,\ell ' \in \Lambda} \cos k_{\ell '} \cos\Theta_{\ell,\ell',S} \nonumber \\ 
& + \sqrt{ \sum_{\ell,\ell ',\ell '',\ell ''' \in \Lambda} \!\!\cos k_{\ell '}\cos k_{\ell '''} 
\cos \big( \Theta_{\ell,\ell',S}-\Theta_{\ell'',\ell''',S} \big) } \Bigg\} .
\end{align}
where we defined the angle $\Theta_{\ell,\ell',S}=(k_{\ell} - k_{\ell '})S +(\phi_{\ell}-\phi_{\ell '})$.
The other possible value of the ergotropy~$\mathcal{E}_S^{<}$ is
\begin{eqnarray}
\mathcal{E}_S^{<} & = & 2 |\Delta|\Biggl\lvert \dfrac{2\mathcal{N}^2}{L}\Big\lvert\sum_{\ell \in \Lambda}e^{i(k_{\ell} S + \phi_{\ell})}\Big\rvert^2  - 1 \Biggr\rvert \nonumber \\
& &  + \dfrac{8 J \mathcal{N}^2}{L} \sum_{\ell,\ell ' \in \Lambda} \cos k_{\ell '} \cos\Theta_{\ell,\ell',S}.
\end{eqnarray}

We now specialize to the simple case of two superposed current states: $\ket{\Psi^{(2)}}=\frac{1}{\sqrt{2}}\left(\ket{\ell_1} + e^{i\phi_{21}}\ket{\ell_2}\right)$, for which $f_j =\frac{1}{\sqrt{2}} \big( e^{ik_1 j} + e^{i(k_2 j + \phi_{21})}\big)$, $k_1=2\pi\ell_1/L$, and $k_2=2\pi\ell_2/L$. As shown before, the computation of the LE passes through the correlation functions and the onsite magnetization, which read
\begin{subequations}
\label{eq:CM_two}
\begin{eqnarray}
    C_S^{xx} & = & \dfrac{2}{L}\left(\cos k_1 + \cos k_2 \right) \left( 1 + \cos \Phi_{12} \right), \\
    C_S^{xy} & = & \dfrac{2}{L}\left(\cos k_2 - \cos k_1 \right) \sin \Phi_{12},
    \label{eq_Cxy_two} \\
    M_S^z & = & \dfrac{2}{L} \big( 1 + \cos\Phi_{12} \big) - 1,
\end{eqnarray}
\end{subequations}
where we defined $\Phi_{12} = (k_1 - k_2)S - \phi_{21}$.
Substituting them in Eqs.~\eqref{eq:E_S<>}, we finally obtain
\begin{eqnarray}
\mathcal{E}_S^{>} & \! = \! & \dfrac{4J}{L} \bigg\{ \Big[ (\cos k_1 + \cos k_2)^2 (1 + \cos \Phi_{12}^2)^2 \nonumber \\ && \qquad \quad + (\cos k_1 - \cos k_2)^2 \sin^2 \Phi_{12} \Big]^{1/2} \qquad \nonumber\\ &&
\qquad + (\cos k_1 + \cos k_2) (1 + \cos \Phi_{12}) \bigg\},
\label{eq:ES>_twoell} \\
\mathcal{E}_S^{<} & \! = \! & 2|\Delta|\Bigl\lvert \dfrac{2}{L}\left( 1 + \cos \Phi_{12} \right) - 1 \Bigr\rvert \nonumber \\ && + \dfrac{4J}{L} (\cos k_1 + \cos k_2) (1 + \cos \Phi_{12}).
\label{eq:ES<_twoell}
\end{eqnarray}

\section{Energy per site of a superposition of two current states}
\label{app:per_site_energy}

In this section, we discuss the behavior of the energy per site of a superposition state 
\begin{equation}
    \ket{\Psi^{(2)}}=\tfrac{1}{\sqrt{2}}(\ket{\ell_1} + \ket{\ell_2}).
\end{equation}
We first rewrite the system Hamiltonian in Eq.~\eqref{eq:XY_NN_ham} as 
\begin{equation}
 \hat{H} = \frac{1}{2} \sum_j (\hat{h}_{j-1,j} + \hat{h}_{j,j+1}) + \Delta \sum_j \hat{\sigma}^z_j , 
\end{equation}
where $\hat{h}_{j,j+1} = J \big( \hat{\sigma}_j^x\hat{\sigma}_{j+1}^x + \hat{\sigma}_j^y\hat{\sigma}_{j+1}^y \big)$. 
Identifying $\hat{\epsilon}_j = \frac{1}{2}  (\hat{h}_{j-1,j} + \hat{h}_{j,j+1})$ as the energy contribution to the $j$th site given by its nearest neighbors, the Hamiltonian reads
$\hat{H}=\sum_j (\hat{\epsilon}_j + \Delta\hat{\sigma}^z_j)$. 
We define the mean energy per site of a generic state $\ket{\psi}$ as
\begin{equation}
    E_j [\ket{\psi}] = \braket{\psi|\hat{\epsilon}_j + \Delta\hat{\sigma}^z_j|\psi}.
\end{equation}
In the single excitation sector, it reads
\begin{equation}
    E_j [\ket{\psi}] = J C_j^{xx} + \Delta M_j^z. 
\end{equation}
Using Eqs.~\eqref{eq:CM_two}, we compute it for the superposition of two current states and find:
\begin{align}
    E_j\bigl[ \ket{\Psi^{(2)}} \bigr] = & \frac{2J}{L} \big( \cos k_1 + \cos k_2 \big) \big\{1 + \cos \big[ (k_1 - k_2)j \big] \big\} \nonumber \\ &+ \Delta \left\{ \frac{2}{L} + \frac{2}{L}\cos \big[ (k_1 - k_2)j \big] - 1 \right\} ,  
    \label{eq:persite_energy}
\end{align}
for fixed $L$, $\ell_1$ and $\ell_2$.
If we look only at its $j$-dependence, we have
\begin{equation}
    E_j\bigl[ \ket{\Psi^{(2)}} \bigr] \sim \frac{2}{L} \cos \big[ (k_1 - k_2)j \big] \left[ J(\cos k_1 + \cos k_2) + \Delta \right], \nonumber  
\end{equation}
while the rest is a common shift for any site.

We discuss the case of $\cos k_1 + \cos k_2 >0$. For both $J,\Delta < 0$, the shape of energy distribution $E_j$ per site does not change for different values of $|\Delta/J|$, it scales as
$E_j \bigl[ \ket{\Psi^{(2)}} \bigr] \sim - \cos[(k_1 - k_2)j] = -\cos [ 2\pi(\ell_1 -\ell_2) j/L ]$. For the specific choice of $|k_1-k_2|=2\pi/L$, the energy per site is peaked in $j=L/2$ for even $L$ and in $j=(L\pm 1)/2$ for odd $L$, it presents a minimum in $j=L$. 

\begin{table}[h]
\centering
\begin{tabular}{||c | c||} 
  \hline
  $L$ & $\cos k_1 + \cos k_2$ \\ 
  \hline \hline
  $10$ & $1.118$  \\ 
  \hline
  $11$ & $1.257$ \\ 
  \hline 
  $12$ & $1.366$  \\ 
  \hline
  $13$ & $1.454$ \\ 
  \hline
  $14$ & $1.524$  \\ 
  \hline
  $15$ & $1.583$ \\ 
  \hline  
  $16$ & $1.631$  \\ 
  \hline
  $17$ & $1.671$ \\ 
  \hline
  $18$ & $1.706$ \\ [1ex]
  \hline
\end{tabular}
\caption{Values of $|\Delta|/J$ for which the LE distribution changes shape, for different values of $L$, $\Delta$ negative and $J$ positive.}
\label{table:change_shape}
\end{table}
 For positive $J$, and negative $\Delta$ the energy per site scales as
\begin{equation}
E_j \bigl[ \ket{\Psi^{(2)}} \bigr] \sim 
   \begin{cases}
    - \cos[(k_1 - k_2)j] & \textrm{if} \; \; \frac{|\Delta|}{J } >\cos k_1 \!+\! \cos k_2 \\
 \phantom{-} \cos[(k_1 - k_2)j] & \textrm{if} \; \; \frac{|\Delta|}{J} < \cos k_1 \!+\! \cos k_2.   
    \end{cases}
\end{equation}
Thus, for $|\Delta| / J > \cos k_1 + \cos k_2 $ the distribution has the minimum in $j=L$ and it is peaked in $j=L/2$ for even $L$ and in $j=(L\pm 1)/2$ for odd $L$. Conversely, for $|\Delta| / J<\cos k_1 + \cos k_2 $, the distribution is peaked in $j=L$ and is minimum in $j=L/2$ for even $L$ and in $j=(L\pm 1)/2$ for odd $L$. There is a competition between the spin exchange $JC_j^{xx}$ and the field term $\Delta M_j^z$ contributions to the energy per site which is responsible for a change of the shape of the distribution of the latter. In Table~\ref{table:change_shape}, we fix $\ell_1=1$, $\ell_2 = 2$ and we report the particular value of $|\Delta|/J=\cos k_1 + \cos k_2$ for which the shape of the energy per site distribution changes shape.

The specific values for which the energy per site changes shape are the same as those for which the LE distribution changes shape, reported in Fig.~\ref{fig:ergdistr}(b).
\section{Comparison between single current states and quantum superposition}
\label{app:comparison}
We compare the maximum over all sites of the LE of a superposition of two current states with the LE of the single current state. Using Eqs.~\eqref{eq:Jnega},\eqref{eq:CSXX},\eqref{eq:CS_XY},\eqref{eq:MSZ}, it is possible to obtain the LE of a single current state $\ket{\ell}$ for both $J\lessgtr 0$:
\begin{align}
\dfrac{\mathcal{E}_S}{|J|}=
\begin{cases}
\dfrac{8}{L}\left[\left|\cos\left(\frac{2\pi\ell}{L} \right)\right| +\textrm{sgn}(J) \cos\left(\frac{2\pi\ell}{L} \right)\right]\; \; \textrm{if} \; \; \bigg| \dfrac{\Delta}{J} \bigg| < g^{(\ell)} , \\ \\
\dfrac{2}{L} \left[ \bigg| \dfrac{\Delta}{J} \bigg|(L-2) + 4\textrm{sgn}(J) \cos\left(\frac{2\pi\ell}{L} \right) \right]\; \;  \textrm{if} \; \;  \bigg| \dfrac{\Delta}{J} \bigg| > g^{(\ell)},
\end{cases}
\label{eq:Jposneg_ell}   
\end{align}
where $g^{(\ell)}\coloneq 4 |\cos(2\pi\ell/L)| /(L-2)$. We observe that the LE changes behavior for $(L-2)|\Delta|\lessgtr 4\bigl|J \cos(2\pi\ell/L)\bigr|$, meaning $|E_{\Delta}|\lessgtr |E_{\ell}|$. $E_{\ell}=4J \cos(2\pi\ell/L)$ is the spin exchange energy $E_{\rm hop}$ associated to the state $\ket{\ell}$. $E_{\Delta}$ and $E_{\rm hop}$ are defined in Appendix~\ref{app:XZ_extractable} and are the energies associated respectively to the detuning and spin exchange terms of the Hamiltonian. We assume positive $\cos(2\pi\ell/L)$, for positive $J$ and dominating spin exchange energy, the LE is $4E_{\ell}/L$. Otherwise, for dominating detuning energies it scales as $|E_{\Delta}|$ plus positive corrections. For negative $J$ and dominating spin exchange energy, the LE is zero, for dominating detuning energy, it scales as $|E_{\Delta}|$ plus negative corrections. 

Figure~\ref{fig:maximum_ergotropy} of the main text reports the behavior of the LE of two current states $\ket{\ell_1=1}$, $\ket{\ell_2=2}$ compared with the maximum LE over the ring of their superposition. The ring is composed of $L=11$ sites, thus in both cases $\cos(2\pi\ell_a/L)>0$ ($a=1,2$). Both cases of $J$ positive (dashed lines) and $J$ negative (solid lines) are considered. We first discuss the case $J>0$. For $|\Delta/J|\ll 1$, the LE of the superposition state is given by Eq.~\eqref{eq:ES>_twoell} and it is peaked in $S=L$ for our particular choice of the parameters. From Eq.~\eqref{eq:ES>_twoell}, it is possible to show that the maximum value of the LE of the superposition state $\mathcal{E}_{S=L}^{(\ell_1,\ell_2)}$ is the sum of those of the two superposed states, meaning $\mathcal{E}_{S=L}^{(\ell_1,\ell_2)}=4[E_{\ell_1} + E_{\ell_2}]/L$.  Conversely, for $|\Delta/J|\gg 1$, the LE is peaked in $S=(L\pm 1)/2$ (even $L$) and is given by Eq.~\eqref{eq:ES<_twoell}. It scales as $2|\Delta||1-\frac{2}{L}(1-\cos(\pi/L))|\approx 2|\Delta||1-\frac{\pi^2}{L^3}|$ for sufficiently big $L$. In this limit, the LE of the single current state scales as $\frac{2(L-2)}{L}|\Delta|=2|\Delta|\left(1-\frac{2}{L}\right)$ and so it has a smaller slope of those of the superposition. For intermediate values of $|\Delta/J|$, the LE of the superposition state is not always dominant (see Fig.~\ref{fig:maximum_ergotropy}). Regarding the case $J<0$, at $|\Delta/J|=0$, the LE of all the states considered is zero. While for the single current states, it starts to grow linearly after a certain threshold of $|\Delta/J|$, for the superposition of two current states, the LE grows linearly from $|\Delta/J|=0^+$ with fixed slope. For the same reason as those discussed in the case $J>0$, for $|\Delta/J|\gg 1$ the LE of the superposition state grows faster in $|\Delta|$ than those of the single current states. In general, the LE of the superposition state remains bigger than those of the single current superposed states for any value of $|\Delta/J|$.

\section{Extractable energy through $X$ and $Z$ transformations}
\label{app:XZ_extractable}

\begin{figure}[h]
\centering
\includegraphics[width=0.9\linewidth]{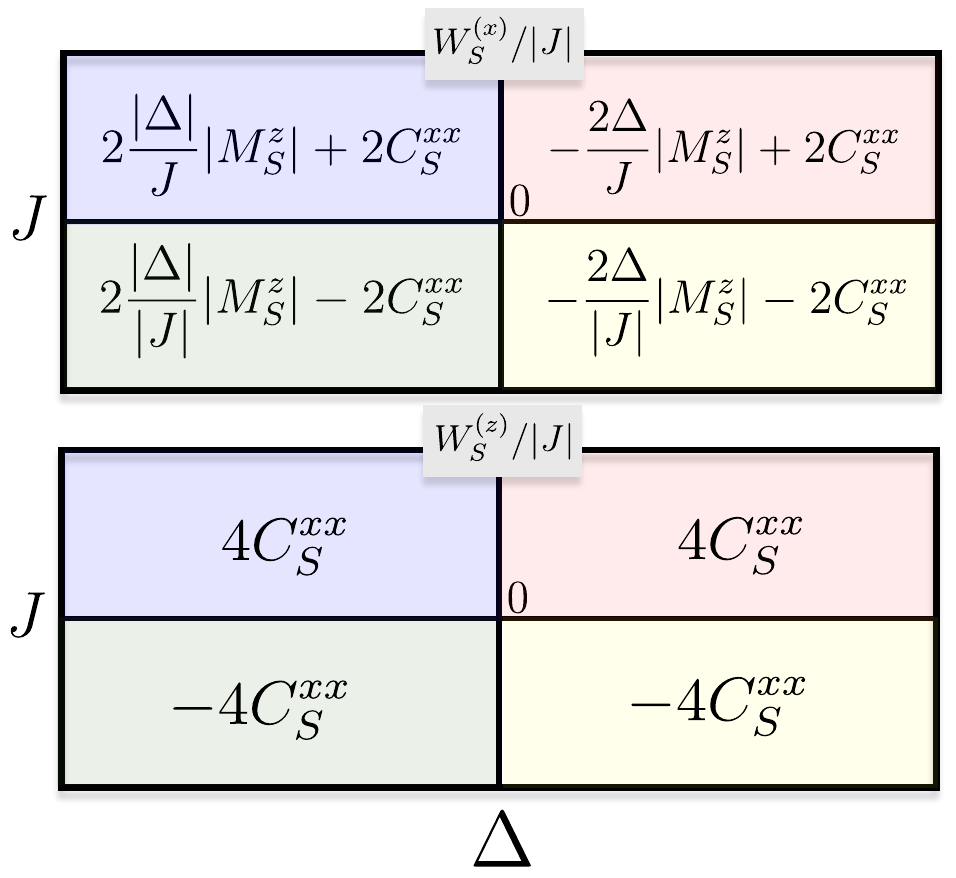}
\caption{Schematic representation of the behavior of the extractable energies through $X$ (top panel) and $Z$ (bottom panel) transformations in the $(\Delta,J)$ plane. A state with negative $z$-magnetization on each site is considered.}
\label{fig:Wzx_scheme}
\end{figure}

\begin{figure*}[!t]
\centering
\includegraphics[width=\textwidth]{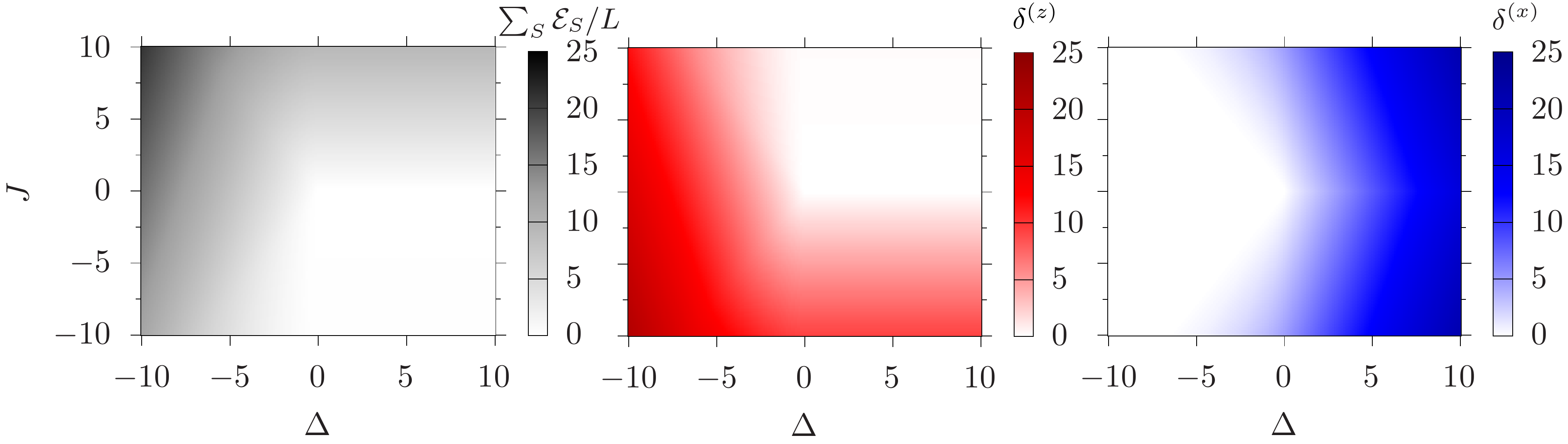}
\caption{LE averaged over the ring, $\delta^{(z)}$ and $\delta^{(x)}$ in function of spin exchange strength $J$ and field $\Delta$. The number of sites is $L=11$, $\ell_1=1$ and $\ell_2=2$ are the two superposed states.}
\label{fig:Errors}
\end{figure*}

Here, we compute the extractable energy through local $\hat{\sigma}^x_S$ and $\hat{\sigma}^z_S$ transformations. We consider a generic one-excitation state of Eq.~\eqref{eq:psi_1s}. Its spin flipped $\ket{\psi^x_S}$ state is given by 
\begin{equation}
    \ket{\psi^x_S} \equiv \hat \sigma^x_S \ket{\psi_{\rm 1s}} = \sum_j f_j \big( \hat{\sigma}_S^- + \hat{\sigma}_S^+ \big) \hat{\sigma}_j^+
\ket{0},
\end{equation}
where we used $\hat{\sigma}_S^x = \hat{\sigma}_S^+ + \hat{\sigma}_S^-$ and $\ket{0} \equiv \ket{\downarrow, \ldots, \downarrow}$. We write the Hamiltonian~\eqref{eq:XY_NN_ham} as $\hat{H} = \hat{H}_{\rm hop} + \hat{H}_{\rm \Delta}$, with
\begin{equation}
\hat{H}_{\rm hop} = \tilde{J} \sum_j \big( \hat{\sigma}_j^+ \hat{\sigma}_{j+1}^- + {\rm H.c.} \big),
\quad
\hat{H}_{\Delta} = \Delta \sum_j \hat{\sigma}_j^z,   
\end{equation}
and $\tilde{J}=2J$. We also define
$E_{\rm hop}[\ket{\phi}] \equiv \braket{\hat{H}_{\rm hop}}_\phi$ and $E_{\Delta}[\ket{\phi}] \equiv \braket{\hat{H}_{\Delta}}_\phi$.
The energy of the non-rotated state $\ket{\psi_{1s}}$ in Eq.~\eqref{eq:psi_1s} is:
\begin{subequations}
\begin{eqnarray}
   E_{\rm hop}[\ket{\psi_{\rm 1s}}] & = &  \tilde{J}\sum_j
  \big( f_j^* f_{j+1} + f_j f_{j+1}^* \big),  
  \label{eq:Ehop} \\
  E_{\Delta}[\ket{\psi_{\rm 1s}}] & = & \Delta (2 - L). 
  \label{eq:Edelta}
\end{eqnarray}
\end{subequations}
To compute the energy of the rotated state $\ket{\psi^x_S}$, we use the relations
\begin{align}
    & \braket{0|\hat{\sigma}_i^-\hat{\sigma}_j^-\hat{\sigma}_k^+\hat{\sigma}_m^-\hat{\sigma}_s^+\hat{\sigma}_l^+|0}=(-2\delta_{s,l} \delta_{m,s} + \delta_{m,s} + \delta_{m,l} ) \times \nonumber\\& \hspace{3cm} \times (-2\delta_{j,k}\delta_{j,l}\delta_{i,l} + \delta_{j,k}\delta_{i,l} + \delta_{i,k}\delta_{j,l}) \nonumber, \\
    & \braket{0|\hat{\sigma}_j^-\hat{\sigma}_s^-\hat{\sigma}_k^z\hat{\sigma}_s^+\hat{\sigma}_{j'}^+|0} = (2\delta_{k,s} + 2\delta_{k,j '} - 1) \times \nonumber \\& \hspace{3cm} \times (-2\delta_{s,j'}\delta_{j,j'} + \delta_{j,j'} + \delta_{j,s}\delta_{s,j'}),  \nonumber
\end{align}
obtaining
\begin{subequations}
\begin{eqnarray}
  E_{\rm hop}[\ket{\psi^x_S}] & = & \tilde{J}\sum_j
\big( f_j^* f_{j+1} + f_j f_{j+1}^* \big) \label{eq:Ehop_rot}  \\
& & - \tilde{J}f_S (f_{S-1}^* + f_{S+1}^*) - \tilde{J}f_S^* (f_{S-1} + f_{S+1}), \nonumber \\
    E_{\Delta}[\ket{\psi^x_S}] & = & 
    \Delta (4 - L - 4 |f_S|^2).
    \label{eq:Edelta_rot}
\end{eqnarray}
\end{subequations}
We can write the energy of the rotated state in terms of those of the non-rotated substituting Eqs.~\eqref{eq:Ehop},\eqref{eq:Edelta},\eqref{eq:CSXX},\eqref{eq:MSZ} into Eqs.~\eqref{eq:Ehop_rot},\eqref{eq:Edelta_rot}, we obtain
\begin{subequations}
\begin{eqnarray}
    E_{\rm hop}[\ket{\psi^x_S}] & = & E_{\rm hop}[\ket{\psi_{\rm 1s}}] - \tilde{J}\braket{\hat{\sigma}_S^x \otimes \hat{X}_E}_{\psi_{\rm 1s}}, \\
    E_{\Delta}[\ket{\psi^x_S}] & = & E_{\Delta}[\ket{\psi_{\rm 1s}}] - 2 \Delta \braket{\hat{\sigma}_S^z}_{\psi_{\rm 1s}}.
\end{eqnarray}
\end{subequations}
Therefore, given the total energy $E=E_{\rm hop} + E_{\Delta}$, the extractable work $W_S^{(x)}$ though spin-flip operations is
\begin{equation}
W_S^{(x)} \! = \! E[\ket{\psi_{\rm 1s}}]-E[\ket{\psi^x_S}] \! = \! 2J\braket{\hat{\sigma}_S^x \otimes \hat{X}_E}_{\psi_{\rm 1s}} + 2 \Delta \braket{\hat{\sigma}_S^z}_{\psi_{\rm 1s}}
\end{equation}
where we used $\tilde{J}=2J$. 

Using the same procedure, we can compute the energy of the state 
\begin{equation}
\ket{\psi^z_S} \equiv \hat \sigma^z_S \ket{\psi_{\rm 1s}} = \sum_j \hat{\sigma}_S^z\hat{\sigma}_j^+\ket{0}.
\end{equation}
To do so, we use the relation
\begin{equation}
\braket{0|\hat{\sigma}_k^-\hat{\sigma}_s^z\hat{\sigma}_i^+\hat{\sigma}_j^-\hat{\sigma}_s^z\hat{\sigma}_m^+|0}=(2\delta_{s,m} -1)\delta_{k,i}\delta_{j,m}(2\delta_{s,i} - 1)   
\nonumber
\end{equation}
and we get
\begin{subequations}
\begin{eqnarray}
    E_{\rm hop}[\ket{\psi^z_S}] & = & E_{\rm hop}[\ket{\psi_{\rm 1s}}] - 4 J \braket{\hat{\sigma}_S^x \otimes \hat{X}_E}_{\psi_{\rm 1s}},\\
    E_{\Delta}[\ket{\psi^z_S}] & = & E_{\Delta}[\ket{\psi_{\rm 1s}}] \, ,    
\end{eqnarray}
\end{subequations}
thus the extractable work is
\begin{equation}
W_S^{(z)}= E[\ket{\psi_{\rm 1s}}]-E[\ket{\psi^z_S}]=4J\braket{\hat{\sigma}_S^x \otimes \hat{X}_E}_{\psi_{\rm 1s}}.
\end{equation}
To summarize, we obtained 
\begin{equation}
W_S^{(x)} = 2 (J C_S^{xx} + \Delta M_S^z),
\quad
W_S^{(z)} = 4 J C_S^{xx}.
\label{eq:WsXZ}
\end{equation}
Figure~\ref{fig:Wzx_scheme} sketches the behavior of the dimensionless extractable work $W_S^{(\alpha)}/|J|$ in the $(\Delta,J)$ plane, $\alpha=x,z$. We consider a state with negative $z$-magnetization on each site, for instance a superposition of two current states with $L>4$.
To give a quantitative picture of which are the optimal and quasi-optimal transformations, we compute the discrepancy between ergotropy and extracted energy $\delta_S^{(\alpha)} = \mathcal{E}_S - W_S^{(\alpha)}$~\eqref{eq:distance} and we take its average over the ring
\begin{equation}
    \delta^{(\alpha)}=\frac{1}{L}\sum_S \delta_S^{(\alpha)},
\end{equation}
where $W_S^{(\alpha)} = E[\ket{\psi}] - E[\hat{\sigma}_S^{\alpha}\ket{\psi}]$ is the energy extracted through a local $\hat{\sigma}_S^{\alpha}$ transformation.

In Figure~\ref{fig:Errors} we show the behavior of the averaged discrepancies and LE for a specific choice of the superposition state. The plot gives a visual idea of what are the optimal local transformations in the various regions of the $(\Delta,J)$ plane. The region $\Delta>0,J<0$ is characterized by both $\delta^{(x)}$ and $\delta^{(z)}$ non zero and small total LE. On the other hand, the region $\Delta,J>0$ is characterized by small $\delta^{(z)}$, since the $Z$ transformation is quasi optimal for any value of $\Delta$. For $\Delta<0$, if $|\Delta/J|>g_s$ holds on all the sites, the $X$ transformation is optimal everywhere; for this reason, we observe a triangular white region in the $\delta^{(x)}$ plot. The border of the aforementioned triangle is broad, since decreasing $|\Delta|$ the number of sites in which $X$ optimal decreases; conversely, the number of sites in which $Z$ is quasi optimal increases.

\section{Derivation of the $\mathcal{M}$ matrix for the long-range XY model}
\label{sec:M_XXmodel}

We now consider the long-range XY model of Eq.~\eqref{eq:Ham_dipolar}.
Truncating all the hoppings with a range more than nearest-neighbor, we recover the Hamiltonian of Eq.~\eqref{eq:XY_NN_ham},
with $J = 2g/\big[ 2R\sin(\pi/L) \big]^{\alpha}$ 
denoting the nearest-neighbor hopping strength.
As the subsystem $S$, we take a single site of the chain, while the environment $E$ is all the rest. The subsystem-environment interaction can be written as
\begin{equation}
\hat{V}_{SE} = \sum_{j \neq S} J_{Sj} \big( \hat{\sigma}_S^x \hat{\sigma}_j^x + \hat{\sigma}_S^y \hat{\sigma}_j^y \big) .
\end{equation}
Since $\hat{H}_S$ is the same as for the nearest-neighbor model [cf.~Eq.~\eqref{eq:Ham_S}], the $\mathcal{M}_{xx}$ matrix element is
\begin{equation}
\mathcal{M}_{xx}=-\frac{1}{2}\textrm{Tr}_E \left[ \Tr_S\{\hat{\sigma}_S^x\rho_{SE}\}  \Tr_S\{\hat{\sigma}_S^{x}\hat{V}_{SE}\}\right],
\end{equation}
where we used $\textrm{Tr}_S\{\hat{\sigma}_S^{x}\rho_S\}=0$. We observe that
\begin{align}
\Tr_S \{\hat{\sigma}_S^{x}\hat{V}_{SE}\} &= \sum_{j \neq S} J_{Sj} \textrm{Tr}_S\left[ \hat{\sigma}_S^x \big( \hat{\sigma}_S^x\hat{\sigma}_j^x +  \hat{\sigma}_S^y\hat{\sigma}_j^y \big) \right]\nonumber\\&= \sum_{j \neq S} J_{Sj} \Big( \textrm{Tr}_S \{ \hat{\mathbb{1}}_S \}\hat{\sigma}_j^x + i \textrm{Tr}_S \{ \hat{\sigma}_S^z \} \hat{\sigma}_j^y \Big) \nonumber \\
& = 2 \sum_{j \neq S} J_{Sj} \big( \hat{\sigma}_j^x \big)_E
\end{align}
where $\big( \hat{\sigma}_j^x \big)_E$ is the operator acting as $\hat{\sigma}^x$ on $j$ and as the identity everywhere else in $\mathcal{H}_E$. 
Thus, the $\mathcal{M}_{xx}$ matrix element reads
\begin{equation}
\mathcal{M}_{xx} = -\sum_{j \neq S} J_{Sj} \textrm{Tr}_E \left[ \big( \hat{\sigma}_j^x \big)_E \Tr_S\{\hat{\sigma}_S^x\rho_{SE}\} \right] ,
\end{equation}
which can be rewritten as
\begin{equation}
\mathcal{M}_{xx}=-\sum_{j \neq S} J_{Sj} \braket{\hat{\sigma}_S^x\hat{\sigma}_{j}^x}_{\psi}.  
\label{eq:MXXlong}
\end{equation}

Using the same procedure, we can compute the other matrix elements $\mathcal{M}_{yy}$, $\mathcal{M}_{xy}$, and $\mathcal{M}_{yx}$
[cf.~Eq.~\eqref{eq:Mdef}].
Since $\hat{H}_S$ is unchanged and $\hat{V}_{SE}$ still contains only $\hat{\sigma}^x$ and $\hat{\sigma}^y$ terms, the $\mathcal{M}_{zz}$ component is the same as the one in Eq.~\eqref{eq:Mzz_nn}, which was obtained for the nearest-neighbor model.
The other components of the $\mathcal{M}$ matrix,  $\mathcal{M}_{zx}=-\sum_{j \neq S} J_{Sj} \braket{\hat{\sigma}_S^z\hat{\sigma}_{j}^x}_{\psi}$, $\mathcal{M}_{zy}=-\sum_{j \neq S} J_{Sj} \braket{\hat{\sigma}_S^z\hat{\sigma}_{j}^y}_{\psi}$, $\mathcal{M}_{xz}=-\Delta \braket{\hat{\sigma}_S^z}_{\psi}$ and $\mathcal{M}_{yz}=-\Delta \braket{\hat{\sigma}_S^y}_{\psi}$, are all equal to zero in the one-excitation sector, i.e., for $\ket{\psi} = \ket{\psi_{\rm 1s}}$, analogously as for the nearest-neighbor model.

Summarizing, we obtain the following expressions for the non-zero components of the $\mathcal{M}$ matrix evaluated on the state in Eq.~\eqref{eq:psi_1s}:
\begin{subequations}
\begin{align}
&\mathcal{M}_{xx}=-\sum_{j \neq S} J_{Sj} (f_S^* f_j + f_j^* f_S) = \mathcal{M}_{yy}, \\
&\mathcal{M}_{xy}=-i\sum_{j \neq S} J_{Sj} (f_S^* f_j - f_j^* f_S) = - \mathcal{M}_{yx}, \\
&\mathcal{M}_{zz} = \Delta (1 - 2|f_S|^2).
\end{align}
\end{subequations}
Note that, disregarding hoppings from the next-to-nearest-neighbor to all the longer-range ones, we recover the matrix elements found for the nearest-neighbor model in Eq.~\eqref{eq:XY_NN_ham} (see the Appendix~\ref{app:M_XYmodelNN}).

\end{document}